\documentclass[aps, prl, reprint, amsmath, amssymb, groupedaddress]{revtex4-1}

\usepackage{graphicx, subfigure, dcolumn}
\usepackage{bm}

\begin{document}

\title{Universal Continuous Variable Quantum Computation Without Cooling}

\author{Hoi-Kwan Lau}\email[Email address: ]{hklau.physics@gmail.com}
\affiliation{Institute of Theoretical Physics, Ulm University, Albert-Einstein-Allee 11, 89069 Ulm, Germany}
\author{Martin B. Plenio}
\affiliation{Institute of Theoretical Physics, Ulm University, Albert-Einstein-Allee 11, 89069 Ulm, Germany}

\date{\today} 

\begin{abstract}
One of the limitations to the quantum computing capability of a continuous-variable system is determined by our ability to cool it to the ground state, because pure logical states, in which we accurately encode quantum information, are conventionally pure physical states that are constructed from the ground state.
In this work, we present an alternative quantum computing formalism that encodes logical quantum information in \textit{mixed} physical states.  
We introduce a class of mixed-state protocols that are based on a parity encoding, and propose an implementation of the universal logic gates by using realistic hybrid interactions.  When comparing with the conventional pure-state protocols, 
our formalism could relax the necessity of, and hence the systemic requirements of cooling.
Additionally, the mixed-state protocols are inherently resilient to a wider class of noise processes, and reduce the fundamental energy consumption in initialisation.
Our work broadens the candidates of continuous-variable quantum computers.  
\end{abstract}


\maketitle

\textit{Introduction--}  Quantum computers are expected to outperform classical computers in a wide class of applications such as factoring large numbers, database searching, and simulating quantum systems \cite{book:NielsenChuang}.  
The basic logical unit of quantum information is usually a two-level system that can be prepared in an arbitrary superposition state (qubit).  A suitable platform for implementing quantum computers should exhibit `well characterised' physical states for representing the qubit bases $|0_L\rangle$ and $|1_L\rangle$ \cite{DiVincenzo:2000vw}.  In continuous-variable (CV) quantum systems, such as cavity modes of electromagnetic wave, mechanical oscillators, and spin ensembles \cite{Braunstein:2005wr,Poot:2012fh,Haffner:2008tg,Tordrup:2008kf,Wesenberg:2009es}, there is no standard representation of a qubit because each physical degree of freedom (qumode) consists of an abundance of evenly-spaced energy eigenstates.  
Different CV encodings have been invented to represent a qubit as a superposition of Fock states, coherent states, cat states, superpositions of squeezed states, and else \cite{Chuang:1995ud, Chuang:1997ko, Knill:2001vi, Ralph:2003jk, Anonymous:hr, Leghtas:2013ff,Mirrahimi:2014js, Gottesman:2001vk, Menicucci:2014cx, Chuang:1997dx, Ketterer:2015Code, Michael:2016code}.

Despite their differences in detail, all the existing encodings (pure-state encodings) commonly require a pure logical state qubit to be represented by a pure physical state,
which is usually prepared from the ground state of a qumode.  If the qumode frequency is high, a near-ground state with negligible thermal excitation can be obtained by lowering the background temperature \cite{OConnell:2010br}.  Whereas a low frequency qumode has to be cooled by additional processes, e.g., using feedback controls or coupling to dissipative auxiliary systems \cite{Cirac:1992tp, Marzoli:1994tz, Hopkins:2003hy, Poot:2012fh, WilsonRae:2007jp,Marquardt:2007dn, Jaehne:2008gw, Teufel:2011jg, Chan:2011dy,Machnes:2010bg, Machnes:2012bg, Martin:2004kv, WilsonRae:2004il, Xia:2009kl, Rabl:2010cm, Kepesidis:2013fv, Lau:2016cool}.
Despite some remarkable success, ground-state cooling remains challenging for some CV systems due to the lack of appropriate internal structure, absorption heating induced by the cooling laser, and other limitations of apparatus and system.
The ability of achieving ground-state cooling is therefore deemed an important criterion for discriminating CV candidates of a quantum computer.

Nevertheless, quantum computers could be implemented in a wider range of CV systems if the requirements of pure physical states, and thus ground-state cooling, are relaxed.
In fact, there are three pieces of evidence in the literature that hint at CV mixed-state universal quantum computation (MSUQC).  In Refs.~\cite{Jeong:2006gs, Jeong:2007bc}, Jeong and Ralph discovered a macroscopic quantum property that can be observed from the superposition of displaced thermal states.
Although such quantumness can be applied in quantum teleportation and verifying Bell's inequality, the implementation of universal logic gates is unknown. 
In Ref.~\cite{Grace:2006gc}, Grace \textit{et.~al.} discussed the general criteria for MSUQC.  
However, deriving an explicit CV protocol is not obvious based on their results, 
and we will show later that the example considered in this reference is only trivially mixed.
In Ref.~\cite{Lau:2016UUQC}, Lau and Plenio proposed two explicit CV protocols of universal quantum computation that require only partial knowledge about the encoding states.  However, these protocols require the preparation of copies of states, so they cannot be implemented with thermal states which are uncorrelated among qumodes.  

In this work, we affirm the possibility of CV MSUQC by designing a class of explicit protocols 
that pure logical qubits are encoded in mixed-state qumodes.  We note that when comparing to early mixed-state quantum computers that are not scalable, such as the nuclear magnetic resonance implementations \cite{Gershenfeld:1997ie}, and mixed-state applications that perform only specific tasks, like the deterministic quantum computation with one pure state \cite{Knill:1998us,PhysRevA.93.052304}, our protocols are both scalable and universal.  

\textit{Mixed-State Quantum Computation--}  
In pure-state universal quantum computation, a pair of orthogonal physical states, $\mathbb{Q}=\{ |\psi_0\rangle, |\psi_1\rangle \}$, is first specified as the qubit basis.  These basis states are associated with two logical operators that resemble the algebra of Pauli operators, i.e., $\hat{Z}_L|\psi_i\rangle=(-1)^i|\psi_i\rangle$, 
$\hat{X}_L(|\psi_0\rangle \pm |\psi_1\rangle)=(|\psi_1\rangle \pm |\psi_0\rangle)$, 
and $(\hat{X}_L\hat{Z}_L+\hat{Z}_L\hat{X}_L)|\psi_i\rangle =0$, for both $i=0,1$ \footnote{The $\hat{Y}_L$ operator is defined to be $i\hat{X}_L \hat{Z}_L$ for any $\hat{X}_L$ and $\hat{Z}_L$.}.  
Universal quantum computation is implemented by the physical operations, 
\begin{equation}\label{eq:U_gate}
\hat{U}_X(\theta)=e^{i\theta \hat{X}_L},~\hat{U}_Z(\theta)=e^{i\theta \hat{Z}_L},~\hat{U}_{ZZ}(\theta)=e^{i\theta \hat{Z}_L\otimes \hat{Z}_L}.\\
\end{equation}
The first two operations constitute arbitrary angle single logical qubit rotation, and the last operation is the logical entanglement gate.  The computational result is readout by a $\hat{Z}_L$ measurement that distinguishes $|\psi_0\rangle$ from $|\psi_1\rangle$.
 
Instead of restricting to only one basis pair, MSUQC allows multiple pairs of basis states, $\mathbb{Q}_n=\{|\psi_0^{(n)}\rangle,|\psi_1^{(n)}\rangle\}$, where $\langle\psi_i^{(m)}|\psi_j^{(n)}\rangle=\delta_{mn}\delta_{ij}$  \cite{Grace:2006gc}.  
The mixed-state logical operators are given by
$\hat{Z}_L = \sum_n \hat{Z}^{(n)}_L$ and $\hat{X}_L = \sum_n \hat{X}^{(n)}_L$, where $ \hat{Z}^{(n)}_L=  |\psi_0^{(n)}\rangle \langle\psi_0^{(n)}|-|\psi_1^{(n)}\rangle \langle\psi_1^{(n)}|$ and $\hat{X}^{(n)}_L=|\psi_0^{(n)}\rangle \langle\psi_1^{(n)}| + |\psi_1^{(n)}\rangle \langle\psi_0^{(n)}|$.
The logical operators perform the same Pauli algebra on each basis pair, i.e.,  
$\hat{Z}_L|\psi_i^{(n)}\rangle=(-1)^i|\psi_i^{(n)}\rangle$, 
$\hat{X}_L(|\psi_0^{(n)}\rangle\pm|\psi_1^{(n)}\rangle)=(|\psi_1^{(n)}\rangle\pm|\psi_0^{(n)}\rangle)$, 
and $(\hat{X}_L\hat{Z}_L+\hat{Z}_L\hat{X}_L)|\psi_i^{(n)}\rangle =0$, for any $n$ and $i=0,1$.  
The universal logic gates can be physically implemented by the same way as in Eq.~(\ref{eq:U_gate}).  
The computational result is readout by a $\hat{Z}_L$ measurement that distinguishes $\{|\psi_0^{(n)}\rangle\}$ from $\{|\psi_1^{(n)}\rangle\}$, but the identification of an individual state is not necessary.  

The main idea of MSUQC is that all logical processes are identical for each basis pair, so it is not necessary to know the basis in which quantum information has encoded.  
As an illustration, we consider a general pure-state $K$-qubit computational task whose initial state is $|\Psi_0\rangle \equiv \bigotimes_{k=1}^K|\psi_0\rangle_{L_k}$; the subscript $L_k$ denotes the affiliation to the $k$th logical qubit.  The algorithm is specified by a transformation $\mathcal{U}$, which is a functional of the logical operators of each qubit.  The result of the task is  $\mathcal{A} = |\langle \Psi_0| \mathcal{U}|\Psi_0\rangle |^2$, which is the projective measurement probability of $\bigotimes_{k=1}^K(\hat{\mathbb{I}}_{L_k}+\hat{Z}_{L_k})$, where $\hat{\mathbb{I}}$ is the identity operator.
In MSUQC, the initial state is $\rho_0 = \sum_{\vec{n}} p_{\vec{n}} |\Psi_0^{\vec{n}}\rangle\langle\Psi_0^{\vec{n}}|$, where $\vec{n}=\{n_1,\ldots,n_K \}$; $n_k$ is the basis index of the $k$th qubit; $|\Psi_0^{\vec{n}}\rangle \equiv \bigotimes_{k=1}^K |\psi_0^{(n_k)}\rangle_{L_k}$.  The mixedness comes from the unknown probabilities $p_{\vec{n}}$ that satisfy $\sum_{\vec{n}}p_{\vec{n}}=1$.  
Because the mixed-state logical operations preserve and perform identically on each basis, the final measurement probability is the same for every $\vec{n}$.  Therefore the mixed-state computational result is $ \textrm{Tr}\big\{\mathcal{U} \rho_0 \mathcal{U}^\dag (\sum_{\vec{n}}|\Psi_0^{\vec{n}} \rangle\langle \Psi_0^{\vec{n}}|) \big\}=\sum_{\vec{n}} p_{\vec{n}} |\langle\Psi_0^{\vec{n}}| \mathcal{U} |\Psi_0^{\vec{n}}\rangle|^2 = \sum_{\vec{n}} p_{\vec{n}}\mathcal{A} = \mathcal{A}$, which is identical to the pure-state case.  Detail is given in Appendix I. 
\textit{Parity Encoding and Variants--}  The practical challenge of CV MSUQC is to realise the universal logic gates with existing interactions.  For instance, if each basis pair is formed by consecutive Fock states, $\mathbb{Q}^\textrm{Fock}_n=\{|2n\rangle,|2n+1\rangle \}$,
$\hat{Z}_L$ is then the parity operator of which the exponentiation can be implemented efficiently. 
Unfortunately, we could not find an efficient implementation of $\hat{X}_L$-related operations for this encoding, nor any single-qumode encoding that the universal logic gates can be implemented efficiently.   

Alternatively, we propose a two-qumode parity (TQP) encoding that each basis pair is opposite parity Fock states of two qumodes, i.e., $\mathbb{Q}^\textrm{Fock}_{mn}=\{|2m+1\rangle|2n\rangle, |2n\rangle|2m+1\rangle\}$ for any non-negative integer $m$ and $n$.
For the $k$th logical qubit that is composed of the $(2k-1)$th and $(2k)$th qumodes, the logical operators are $\hat{Z}_{L_k} = \hat{\mathcal{P}}_{2k} $ and $\hat{X}_{L_k}= \hat{S}_{2k-1,2k} $, where the single-qumode parity operator is 
$\hat{\mathcal{P}}_i\equiv\sum_{n=0}^\infty (-1)^n |n\rangle_i \langle n|_i=\exp(i\pi \hat{a}^\dag_i \hat{a}_i)$; 
the swap operator $\hat{S}_{ij}$ is defined by the action $\hat{S}_{ij}\hat{a}_i\hat{S}_{ij}^\dag=\hat{a}_j$ for any $i$ and $j$; $\hat{a}_i$ is the annihilation operator of the $i$th qumode.  
Any state $|\psi\rangle_{L_k}$ in $\mathbb{Q}^\textrm{Fock}_{mn}$ satisfies $\hat{\mathcal{P}}_{2k-1}\hat{\mathcal{P}}_{2k} |\psi\rangle_{L_k} = -|\psi\rangle_{L_k}$.  It is straightforward to check that the algebra of Pauli operators is respected.  

Because the parity and swap operators are Hermitian, in principle the universal logic gates in Eq.~(\ref{eq:U_gate}) can be implemented by applying the interactions with the Hamiltonians $H_Z \propto \hat{\mathcal{P}}$, $H_X \propto \hat{S}$, and $H_{ZZ} \propto \hat{\mathcal{P}}\otimes\hat{\mathcal{P}}$.  To the best of our knowledge, however, these interactions have not been realised in any CV system yet.  Alternatively, the logic gates can be implemented by applying beam-splitter and phase-shift operations, together with the second order hybrid interaction of which the Hamiltonian is $H_2 \propto \hat{Z}_A \hat{a}^\dag \hat{a}$, where $\hat{Z}_A$ is the Pauli $Z$ operator of an auxiliary qubit $A$ (See Appendix II and Refs.~\cite{Lau:2016QML, Lau:2016UUQC, Iman:2016UQE}),
\begin{eqnarray}\label{eq:gate}
\hat{\mathcal{C}}_{2k}\hat{R}_X(\theta) \hat{\mathcal{C}}_{2k} |+\rangle_A |\psi\rangle_{L_k} &=& |+\rangle_A \hat{U}_Z(\theta)|\psi\rangle_{L_k}~,~\nonumber\\
 \hat{\mathcal{B}}^\dag_{k}\hat{\mathcal{C}}_{2k}\hat{R}_X(\theta)\hat{\mathcal{C}}_{2k}\hat{\mathcal{B}}_{k}|+\rangle_A |\psi\rangle_{L_k} &=& |+\rangle_A \hat{U}_X(\theta)|\psi\rangle_{L_k}~,\\
\hat{\mathcal{C}}_{2l}\hat{\mathcal{C}}_{2k}\hat{R}_X(\theta) \hat{\mathcal{C}}_{2k}\hat{\mathcal{C}}_{2l} |+\rangle_A |\psi\rangle_{L_kL_l} &=& |+\rangle_A \hat{U}_{ZZ}(\theta)|\psi\rangle_{L_kL_l} ~.~ \nonumber
\end{eqnarray}
$|\psi\rangle_{L_kL_l}$ is the four-qumode physical state of the $k$th and $l$th logical qubits.  $\hat{\mathcal{C}}_i \equiv \exp(i\frac{\pi}{2} (\hat{\mathbb{I}}_A- \hat{Z}_A) \hat{a}^\dag_{i} \hat{a}_{i})$ is the controlled-parity operator acting on the $i$th qumode; $\hat{\mathcal{B}}_k \equiv \exp(\frac{\pi}{4}(\hat{a}_{2k}\hat{a}^\dag_{2k-1}-\hat{a}_{2k}^\dag\hat{a}_{2k-1}))$ is the 50:50 beam splitter between the two qumodes of the $k$th logical qubit.  The auxiliary qubit state is the same as $|+\rangle_A=(|0\rangle_A+|1\rangle_A)/\sqrt{2}$ before and after all operations; $\hat{R}_\sigma(\theta)\equiv\exp(i\theta \hat{\sigma}_A)$ is the rotation of the auxiliary qubit along the Pauli $\sigma$ axis.

\begin{figure*}
\begin{center}
\includegraphics{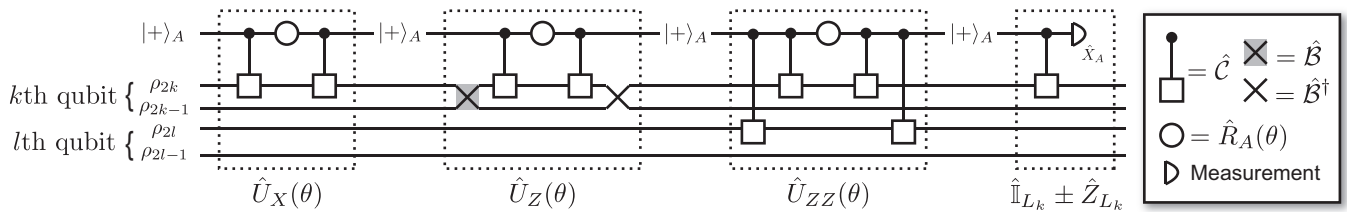}
\caption{ \label{fig:circuit} Circuit diagram of logical processes.  Dotted boxes from left to right: unitary gates $\hat{U}_X(\theta)$, $\hat{U}_Z(\theta)$, $\hat{U}_{ZZ}(\theta)$ in Eq.~(\ref{eq:gate}), and parity measurement in Eq.~(\ref{eq:measure}).  The auxiliary qubit, which remains in $|+\rangle_A$ after each unitary gate, can be reused afterwards.}
\end{center}
\end{figure*}

Measuring the parity of the $(2k)$th qumode is equivalent to a $\hat{Z}_L$ measurement of the $k$th logical qubit.  The parity of any qumode state $|\Psi\rangle$ can be measured destructively \cite{Plick:2010fa}, or by using nondemolition techniques with an auxiliary qubit, i.e., after controlled-parity is applied,
\begin{equation}\label{eq:measure}
\hat{\mathcal{C}}_i |+\rangle_A |\Psi\rangle_i = |+\rangle_A(\hat{\mathbb{I}}_i +\hat{\mathcal{P}}_i)|\Psi\rangle_i/2+|-\rangle_A(\hat{\mathbb{I}}_i -\hat{\mathcal{P}}_i)|\Psi\rangle_i/2~,
\end{equation}
measuring the auxiliary qubit in the Pauli $X$ basis will project $|\Psi\rangle$ to a state with definite parity. 
A circuit diagram of the above logical processes is shown in Fig.~\ref{fig:circuit}. 

The TQP encoding can generate a class of mixed-state encoding that the basis states are $\mathbb{Q}^{\mathcal{V}}_{mn}=\big\{\hat{\mathcal{V}}(|2m+1\rangle|2n\rangle), \hat{\mathcal{V}}(|2n\rangle|2m+1\rangle)\big\}$, where $\hat{\mathcal{V}}$ can be any two-mode unitary.  
In analogy to pure-state encodings, different mixed-state encodings could provide certain benefits due to their differences in the responses to environmental noise and in the required interactions for implementation.  
For instance, $\hat{U}_Z(\theta)$ of $\mathbb{Q}^{\mathcal{V}}_{mn}$ can be implemented in a similar way as Eq.~(\ref{eq:gate}), except that $\hat{\mathcal{C}}$ is replaced by $\hat{\mathcal{V}} \hat{\mathcal{C}} \hat{\mathcal{V}}^\dag = \exp(i\frac{\pi}{2} (\hat{\mathbb{I}}_A- \hat{Z}_A) \hat{\mathcal{V}}\hat{a}^\dag \hat{a} \hat{\mathcal{V}}^\dag)$.  This operation is implemented by the hybrid interaction $H_\mathcal{V}\propto \hat{Z}_A(\hat{\mathcal{V}} \hat{a}^\dag \hat{a}\hat{\mathcal{V}}^\dag)$, instead of $H_2$.  Therefore, engineering an interaction exactly as $H_2$ is not necessary, but a wider range of interaction is applicable in implementing MSUQC.  

\textit{Initialisation--}  Before computation, each qumode is in the thermal state, $\rho_\textrm{th} = (1-e^{-\beta})\sum_{n=0}^\infty e^{-n\beta}|n\rangle\langle n|$, where $e^{-\beta}=\frac{\langle n\rangle}{\langle n\rangle+1}$; $\langle n\rangle$ is the mean thermal excitation \cite{book:Ulf}. 
In the initialisation stage, the parity measurement in Eq.~(\ref{eq:measure}) is conducted to project each qumode to either an even parity state, $\rho_\textrm{even} = (1-e^{-2\beta})\sum_{n=0}^\infty e^{-2n\beta}|2n\rangle\langle 2n|$, 
or an odd parity state, $\rho_\textrm{odd} = (1-e^{-2\beta})\sum_{n=0}^\infty e^{-2n\beta}|2n+1\rangle\langle 2n+1|$. 
After rearranging the qumodes, TQP qubits are initialised that each consists of one $\rho_\textrm{even}$ and one $\rho_\textrm{odd}$, i.e., $\rho_0\equiv \rho_\textrm{odd}\otimes\rho_\textrm{even}$, which represents the logical $|0_L\rangle$ in the $\mathbb{Q}^\textrm{Fock}_{mn}$ basis.

\begin{figure}
\begin{center}
\includegraphics{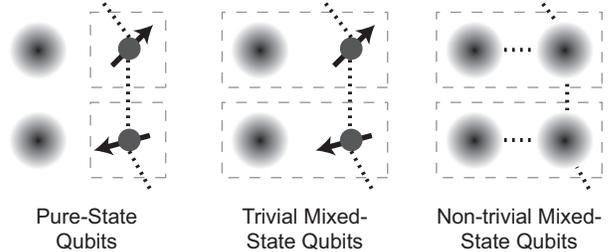}
\caption{ \label{fig:qubit} Left: Conventional quantum computers consist of pure-state qubits (solid circles with arrows) that interact (dotted lines) with each others during computation.  Middle: Each qubit can be viewed as a mixed state if non-interacting mixed-state physical systems (blurred circles) are included in each qubit system (grey dashed boxes).  Right: Non-trivial mixed-state qubits described in this work that consist of multiple interacting physical systems, neither of which is pure.}
\end{center}
\end{figure}

\textit{Triviality of Mixedness--}  
In general, being able to encode pure logical qubits in mixed physical states is not surprising.  
As illustrated in Fig.~\ref{fig:qubit}, 
a pure-state qubit can always be augmented to a mixed-state qubit if non-interacting mixed-state physical degrees of freedom are included in the qubit system. 
The total system entropy of such a mixed-state qubit, which we regard as trivially mixed, 
is equal to the thermal equilibrium entropy of all but one physical degree of freedom.  For example, the mixed-state qubit considered in Ref.~\cite{Grace:2006gc} is a molecule in which the information is encoded in a pure electronic state, while the mixed vibrational state is not involved in the logical processes.  This molecular qubit is trivially mixed because the total system entropy is solely provided by the vibrational degree of freedom.

A two-qumode qubit is trivially mixed only if it consists of a pure-state qumode and an non-interacting thermal state qumode.  The von Neumann entropy of a trivially-mixed two-qumode qubit is, $\mathcal{S}(\rho_\textrm{th})=(\langle n\rangle+1)\log_2(\langle n\rangle+1)- \langle n\rangle \log_2 \langle n \rangle$.  On the other hand, the entropy of a TQP qubit is
\begin{equation}
\mathcal{S}(\rho_0) = 2(\tilde{n}+1)\log_2(\tilde{n}+1)- 2\tilde{n} \log_2 \tilde{n}~,
\end{equation}
where $\tilde{n}=\frac{\langle n\rangle^2}{2\langle n \rangle+1}$.  When $\langle n \rangle \gtrsim 0.8$, we have $\mathcal{S}(\rho_0)>\mathcal{S}(\rho_\textrm{th})$, which shows that the TQP qubit is non-trivially mixed. 

\textit{Implementation Requirements--}  
Because initialising pure physical states is not necessary, MSUQC removes the necessity of ground-state cooling.  Instead, TQP qubits are initialised by nondemolition parity measurement.  Despite certain relationship, nondemolition parity measurement and ground-state cooling are two distinct physical processes that have different implementation requirements \cite{Anders:2010ic}.  In certain situations, TQP encoding could relax the requirements imposed by ground-state cooling.

In some systems, such as the dispersively interacting transmon qubits and cavities \cite{Blais:2004kn}, the qumode and the auxiliary qubit are coupled through the second order hybrid interaction, $H_2 = \lambda \hat{Z}_A \hat{a}^\dag \hat{a}$, where $\lambda$ is the interaction strength.
Because $H_2$ is diagonal in the Fock basis, the qumode cannot be dissipatively cooled even if the auxiliary qubit can be arbitrarily manipulated.  Additional qumode control that is off-diagonal in the Fock basis, such as a displacement operator, is required.  
On the other hand, initialising a TQP qubit by the parity measurement in Eq.~(\ref{eq:measure}) requires only applying $H_2$ for time $t=\pi/2\lambda$, and then projectively measuring the auxiliary qubit.  Therefore TQP encoding relaxes the requirement of off-diagonal qumode control.

\begin{figure}
\begin{center}
\includegraphics{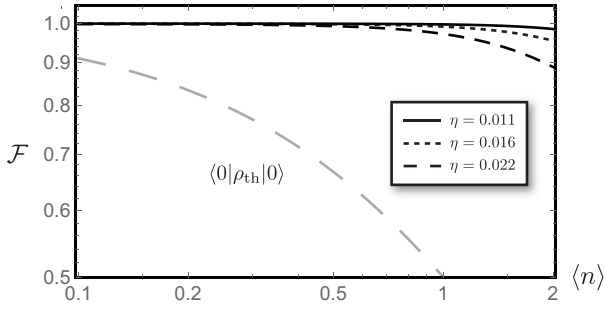}
\caption{ \label{fig:fidelity}
If the implementation of $\hat{\mathcal{C}}$ in Eq.~(\ref{eq:measure}) is inexact, the qumode state becomes $\rho_\pm$ for auxiliary qubit measurement outcome $|\pm\rangle_A$.   Logical qubit fidelity of the state $\rho_-\otimes\rho_+$, which is defined as $\mathcal{F} \equiv \textrm{Tr}\big(\rho_+ (\hat{\mathbb{I}}+\hat{\mathcal{P}})\big)\textrm{Tr}\big(\rho_- (\hat{\mathbb{I}}-\hat{\mathcal{P}})\big)/4$, is shown for initial thermal states with different mean excitation $\langle n\rangle$.  Here $\hat{\mathcal{C}}$ is engineered by repeating the sequence in Eq.~(\ref{eq:H2_engineer}) for 50 times ($\eta=0.022$, dashed), 100 times ($\eta=0.016$, dotted), and 200 times ($\eta=0.011$, solid).  The inaccuracy is mainly induced by the unwanted high order interaction $\propto \eta^4 (\hat{a}^\dag\hat{a})^2$.  Ground state fidelity of the thermal state is also shown for representing pure-state qubit fidelity (grey dashed).}
\end{center}
\end{figure}

For a wider range of CV systems, such as trapped ions \cite{Haffner:2008tg} and mechanical oscillators \cite{Rabl:2009fz}, the prominent hybrid interaction is at the first order, i.e., $H_1 = \eta \nu \hat{\sigma}_A (\hat{a}+\hat{a}^\dag)$, where $\nu$ is the qumode frequency; $\eta$ is the effective Lamb-Dicke parameter  \cite{WilsonRae:2004il}.  The desired operation of $\hat{\mathcal{C}}$ can be engineered by concatenating the dynamics under $H_1$, together with other controls of the qumode and the auxiliary qubit \cite{Lloyd:2003HQC}.  Particularly in the weak coupling regime, $\eta\ll1$, we find a sequence that approximates the dynamics of $H_2$:
\begin{eqnarray}\label{eq:H2_engineer}
&&\left(e^{-i\frac{\pi}{2}\hat{a}^\dag\hat{a}}\hat{D}(- \hat{X}_A 2 \eta)\hat{D}(-\hat{Y}_A 2 \eta)\hat{D}( \hat{X}_A 2 \eta)\hat{D}(\hat{Y}_A 2 \eta)\right)^4 \nonumber \\
&\approx& \exp\big(-i64 \eta^2 \hat{Z}_A (\hat{a}^\dag \hat{a}+1/2)\big) + O(\eta^3)~,
\end{eqnarray}
where $\hat{D}(\alpha)\equiv \exp(\alpha \hat{a}^\dag-\alpha^\ast \hat{a})$ is the displacement operator.   This sequence can be implemented by the free evolutions under $H_1$, and rapidly rotating the auxiliary qubit (details in Appendix III).  The validity of this sequence is numerically verified by simulating the parity measurement in Eq.~(\ref{eq:measure}) (See Fig.~\ref{fig:fidelity}). 

If the sequence (\ref{eq:H2_engineer}) is implemented on a qumode that is interacting with a high temperature background, we show in Appendix IV that the initialisation error of a TQP qubit scales as $\epsilon_\textrm{TQP}\sim \frac{ N_\textrm{th} }{\eta^2 Q} (2\langle n\rangle+1)$, where $Q$ is the qumode quality factor; $N_\textrm{th}$ is the background thermal excitation \cite{Rabl:2010cm}. 

On the other hand, pure-state qubits are most efficiently initialised by resolved-sideband cooling, which requires the engineered decaying rate ($\Gamma_\textrm{dc}$) and dephasing rate ($\Gamma_\textrm{dp}$) of the auxiliary qubit obey, $\Gamma_\textrm{dc},\Gamma_\textrm{dp}\ll \nu$.  Nevertheless, there are systems that the sideband is not resolved due to, e.g. the large $\Gamma_\textrm{dp}$ induced by the cooling process.  For examples, if the qubit dissipation is engineered by a multi-level system, 
$\Gamma_\textrm{dp}$ could be much larger than $\Gamma_\textrm{dc}$ for some intrinsic system parameters \cite{Marzoli:1994tz}.  Another example is if the qumode frequency is low, the frequency linewidth of the classical drive, which rotates the auxiliary qubit, could be broader than the sideband. In these situations that $ \Gamma_\textrm{dc} \ll \nu \ll \Gamma_\textrm{dp} $, the initialisation error of a pure-state qubit is $\epsilon_\textrm{cool}\sim \frac{ N_\textrm{th}}{\eta^2 Q} \frac{\Gamma_\textrm{dp}}{\Gamma_\textrm{dc}}$.  Therefore using the TQP encoding can reduce the initialisation error, i.e., $\epsilon_\textrm{TQP}\ll \epsilon_\textrm{cool}$, when $\langle n \rangle \ll \frac{\Gamma_\textrm{dp}}{\Gamma_\textrm{dc}}$.  


We note that Eq.~(\ref{eq:H2_engineer}) demonstrates only the possibility of implementing TQP encoding with $H_1$, but this sequence is not optimised.  
It is likely that a stronger or more error-resilient $H_2$ can be engineered by, e.g. operating in the strong coupling regime or using another sequence of auxiliary qubit rotations, so $\epsilon_\textrm{TQP}$ could be further reduced. 

\textit{Benefits--}  
Apart from relaxing some system requirements of ground-state cooling, the TQP encoding and its variants reduce the fundamental initialisation energy.  If a thermal state is viewed as a memory of $\mathcal{S}(\rho_\textrm{th})$ bits, ground-state cooling is effectively the erasure of all memory.  According to Landauer's principle, at the background temperature $T$ this process requires a minimum energy consumption of $\mathcal{S}(\rho_\textrm{th}) k_B T \ln 2$, where $k_B$ is the Boltzmann constant \cite{LANDAUER:1961wc, Jacobs:2012eh}.  
Whereas the memory erasure in initialising a TQP qubit is accounted for by the reduction of system entropy, $2\mathcal{S}(\rho_\textrm{th})-\mathcal{S}(\rho_0)$.
The minimum energy consumption is thus $\big(2\mathcal{S}(\rho_\textrm{th})-\mathcal{S}(\rho_0)\big)k_B T \ln 2$, which is smaller than that of ground-state cooling when $\langle n\rangle \gtrsim 0.8$.  

Furthermore, the tensor product structure \cite{Zanardi:2004kb} of a mixed-state qubit can be mathematically expressed as $\rho_{|\Phi\rangle}=|\Phi\rangle \langle \Phi |_\textrm{binary} \otimes \rho_\textrm{other}$ \cite{Grace:2006gc}.  A pure logical qubit $|\Phi\rangle$ is encoded in a binary subsystem, e.g. the two-qumode parity in the TQP encoding, while other degrees of freedom are in mixed states and uninvolved in the computation.  A state with such tensor product structure is inherently a noiseless subsystem  \cite{Knill:2000wt,2000PhRvA..62e2307D,Zanardi:2001vo,Kempe:2001bg} (NS).
A NS-encoded qubit is generally resilient to a wider range of noise than a pure-state qubit.  It is because a NS allows noises to change the physical state except only the encoding subsystem, while a state change implies logical error in pure-state encodings \cite{Zanardi:2001vo,Kribs:2005id}.  
Particularly, we prove in Appendix V that no pure-state encoding can be the decoherence-free-subspace \cite{Lidar:1998wl} of both the collective phase-shift noise, $\hat{\mathcal{E}}_P(\phi)\equiv e^{i\phi\hat{a}^\dag\hat{a}}\otimes e^{i\phi\hat{a}^\dag\hat{a}}$, and the collective squeezing noise, $\hat{\mathcal{E}}_S(\xi) \equiv e^{\xi (\hat{a}^2-\hat{a}^{\dag 2})} \otimes e^{\xi (\hat{a}^2-\hat{a}^{\dag 2})}$.  Whereas the TQP encoding is resilient to both noises, because the logical operators commute with both noise operators, i.e., $[\hat{\mathcal{E}}_P(\phi),\hat{\mathbb{I}} \otimes \hat{\mathcal{P}}] = [\hat{\mathcal{E}}_P(\phi),\hat{S} ]=[\hat{\mathcal{E}}_S(\xi),\hat{\mathbb{I}} \otimes \hat{\mathcal{P}}] = [\hat{\mathcal{E}}_S(\xi),\hat{S} ]=0$.

\textit{Conclusion--}  In this work, we explicitly design a class of universal quantum computation protocols that can be operated with mixed continuous-variable states.  The protocols are derived from an encoding that the qubit information is encoded in the parity of two qumodes.  When comparing with conventional pure-state protocols, the mixed-state protocols relax the necessity and system requirements of cooling, reduce fundamental initialisation energy, and are resilient to a wider class of noise.
Our work provides the possibility for implementing universal quantum computers on the continuous-variable systems where ground-state cooling is challenging.


\textit{Acknowledgement--}  H.-K. L. thanks Christian Weedbrook for useful comments.  This work was supported by the Croucher Foundation, the EU via QUCHIP, and an Alexander von Humboldt Professorship.

\bibliographystyle{apsrev4-1}
\pagestyle{plain}
\bibliography{hUQC_bib}

\begin{thebibliography}{63}%
\makeatletter
\providecommand \@ifxundefined [1]{%
 \@ifx{#1\undefined}
}%
\providecommand \@ifnum [1]{%
 \ifnum #1\expandafter \@firstoftwo
 \else \expandafter \@secondoftwo
 \fi
}%
\providecommand \@ifx [1]{%
 \ifx #1\expandafter \@firstoftwo
 \else \expandafter \@secondoftwo
 \fi
}%
\providecommand \natexlab [1]{#1}%
\providecommand \enquote  [1]{``#1''}%
\providecommand \bibnamefont  [1]{#1}%
\providecommand \bibfnamefont [1]{#1}%
\providecommand \citenamefont [1]{#1}%
\providecommand \href@noop [0]{\@secondoftwo}%
\providecommand \href [0]{\begingroup \@sanitize@url \@href}%
\providecommand \@href[1]{\@@startlink{#1}\@@href}%
\providecommand \@@href[1]{\endgroup#1\@@endlink}%
\providecommand \@sanitize@url [0]{\catcode `\\12\catcode `\$12\catcode
  `\&12\catcode `\#12\catcode `\^12\catcode `\_12\catcode `\%12\relax}%
\providecommand \@@startlink[1]{}%
\providecommand \@@endlink[0]{}%
\providecommand \url  [0]{\begingroup\@sanitize@url \@url }%
\providecommand \@url [1]{\endgroup\@href {#1}{\urlprefix }}%
\providecommand \urlprefix  [0]{URL }%
\providecommand \Eprint [0]{\href }%
\providecommand \doibase [0]{http://dx.doi.org/}%
\providecommand \selectlanguage [0]{\@gobble}%
\providecommand \bibinfo  [0]{\@secondoftwo}%
\providecommand \bibfield  [0]{\@secondoftwo}%
\providecommand \translation [1]{[#1]}%
\providecommand \BibitemOpen [0]{}%
\providecommand \bibitemStop [0]{}%
\providecommand \bibitemNoStop [0]{.\EOS\space}%
\providecommand \EOS [0]{\spacefactor3000\relax}%
\providecommand \BibitemShut  [1]{\csname bibitem#1\endcsname}%
\let\auto@bib@innerbib\@empty
\bibitem [{\citenamefont {Nielsen}\ and\ \citenamefont
  {Chuang}(2010)}]{book:NielsenChuang}%
  \BibitemOpen
  \bibfield  {author} {\bibinfo {author} {\bibfnamefont {M.~A.}\ \bibnamefont
  {Nielsen}}\ and\ \bibinfo {author} {\bibfnamefont {I.~L.}\ \bibnamefont
  {Chuang}},\ }\href@noop {} {\emph {\bibinfo {title} {Quantum Computation and
  Quantum Information}}}\ (\bibinfo  {publisher} {Cambridge University Press},\
  \bibinfo {year} {2010})\BibitemShut {NoStop}%
\bibitem [{\citenamefont {DiVincenzo}(2000)}]{DiVincenzo:2000vw}%
  \BibitemOpen
  \bibfield  {author} {\bibinfo {author} {\bibfnamefont {D.~P.}\ \bibnamefont
  {DiVincenzo}},\ }\href@noop {} {\bibfield  {journal} {\bibinfo  {journal}
  {Fortschritte der Physik}\ }\textbf {\bibinfo {volume} {48}},\ \bibinfo
  {pages} {771} (\bibinfo {year} {2000})}\BibitemShut {NoStop}%
\bibitem [{\citenamefont {Braunstein}\ and\ \citenamefont {van
  Loock}(2005)}]{Braunstein:2005wr}%
  \BibitemOpen
  \bibfield  {author} {\bibinfo {author} {\bibfnamefont {S.~L.}\ \bibnamefont
  {Braunstein}}\ and\ \bibinfo {author} {\bibfnamefont {P.}~\bibnamefont {van
  Loock}},\ }\href@noop {} {\bibfield  {journal} {\bibinfo  {journal} {Reviews
  of Modern Physics}\ }\textbf {\bibinfo {volume} {77}},\ \bibinfo {pages}
  {513} (\bibinfo {year} {2005})}\BibitemShut {NoStop}%
\bibitem [{\citenamefont {Poot}\ and\ \citenamefont {van~der
  Zant}(2012)}]{Poot:2012fh}%
  \BibitemOpen
  \bibfield  {author} {\bibinfo {author} {\bibfnamefont {M.}~\bibnamefont
  {Poot}}\ and\ \bibinfo {author} {\bibfnamefont {H.~S.~J.}\ \bibnamefont
  {van~der Zant}},\ }\href@noop {} {\bibfield  {journal} {\bibinfo  {journal}
  {Physics Reports}\ }\textbf {\bibinfo {volume} {511}},\ \bibinfo {pages}
  {273} (\bibinfo {year} {2012})}\BibitemShut {NoStop}%
\bibitem [{\citenamefont {H{\"a}ffner}\ \emph {et~al.}(2008)\citenamefont
  {H{\"a}ffner}, \citenamefont {Roos},\ and\ \citenamefont
  {Blatt}}]{Haffner:2008tg}%
  \BibitemOpen
  \bibfield  {author} {\bibinfo {author} {\bibfnamefont {H.}~\bibnamefont
  {H{\"a}ffner}}, \bibinfo {author} {\bibfnamefont {C.~F.}\ \bibnamefont
  {Roos}}, \ and\ \bibinfo {author} {\bibfnamefont {R.}~\bibnamefont {Blatt}},\
  }\href@noop {} {\bibfield  {journal} {\bibinfo  {journal} {Physics Reports}\
  }\textbf {\bibinfo {volume} {469}},\ \bibinfo {pages} {155} (\bibinfo {year}
  {2008})}\BibitemShut {NoStop}%
\bibitem [{\citenamefont {Tordrup}\ \emph {et~al.}(2008)\citenamefont
  {Tordrup}, \citenamefont {Negretti},\ and\ \citenamefont
  {M{\o}lmer}}]{Tordrup:2008kf}%
  \BibitemOpen
  \bibfield  {author} {\bibinfo {author} {\bibfnamefont {K.}~\bibnamefont
  {Tordrup}}, \bibinfo {author} {\bibfnamefont {A.}~\bibnamefont {Negretti}}, \
  and\ \bibinfo {author} {\bibfnamefont {K.}~\bibnamefont {M{\o}lmer}},\
  }\href@noop {} {\bibfield  {journal} {\bibinfo  {journal} {Physical Review
  Letters}\ }\textbf {\bibinfo {volume} {101}},\ \bibinfo {pages} {040501}
  (\bibinfo {year} {2008})}\BibitemShut {NoStop}%
\bibitem [{\citenamefont {Wesenberg}\ \emph {et~al.}(2009)\citenamefont
  {Wesenberg}, \citenamefont {Ardavan}, \citenamefont {Briggs}, \citenamefont
  {Morton}, \citenamefont {Schoelkopf}, \citenamefont {Schuster},\ and\
  \citenamefont {M{\o}lmer}}]{Wesenberg:2009es}%
  \BibitemOpen
  \bibfield  {author} {\bibinfo {author} {\bibfnamefont {J.~H.}\ \bibnamefont
  {Wesenberg}}, \bibinfo {author} {\bibfnamefont {A.}~\bibnamefont {Ardavan}},
  \bibinfo {author} {\bibfnamefont {G.~A.~D.}\ \bibnamefont {Briggs}}, \bibinfo
  {author} {\bibfnamefont {J.~J.~L.}\ \bibnamefont {Morton}}, \bibinfo {author}
  {\bibfnamefont {R.~J.}\ \bibnamefont {Schoelkopf}}, \bibinfo {author}
  {\bibfnamefont {D.~I.}\ \bibnamefont {Schuster}}, \ and\ \bibinfo {author}
  {\bibfnamefont {K.}~\bibnamefont {M{\o}lmer}},\ }\href@noop {} {\bibfield
  {journal} {\bibinfo  {journal} {Physical Review Letters}\ }\textbf {\bibinfo
  {volume} {103}},\ \bibinfo {pages} {070502} (\bibinfo {year}
  {2009})}\BibitemShut {NoStop}%
\bibitem [{\citenamefont {Chuang}\ and\ \citenamefont
  {Yamamoto}(1995)}]{Chuang:1995ud}%
  \BibitemOpen
  \bibfield  {author} {\bibinfo {author} {\bibfnamefont {I.~L.}\ \bibnamefont
  {Chuang}}\ and\ \bibinfo {author} {\bibfnamefont {Y.}~\bibnamefont
  {Yamamoto}},\ }\href@noop {} {\bibfield  {journal} {\bibinfo  {journal}
  {Physical Review A}\ }\textbf {\bibinfo {volume} {52}},\ \bibinfo {pages}
  {3489} (\bibinfo {year} {1995})}\BibitemShut {NoStop}%
\bibitem [{\citenamefont {Chuang}\ and\ \citenamefont
  {Yamamoto}(1997)}]{Chuang:1997ko}%
  \BibitemOpen
  \bibfield  {author} {\bibinfo {author} {\bibfnamefont {I.~L.}\ \bibnamefont
  {Chuang}}\ and\ \bibinfo {author} {\bibfnamefont {Y.}~\bibnamefont
  {Yamamoto}},\ }\href@noop {} {\bibfield  {journal} {\bibinfo  {journal}
  {preprint quant-ph/arXiv:9604030}\ } (\bibinfo {year} {1997})}\BibitemShut
  {NoStop}%
\bibitem [{\citenamefont {Knill}\ \emph {et~al.}(2001)\citenamefont {Knill},
  \citenamefont {Laflamme},\ and\ \citenamefont {Milburn}}]{Knill:2001vi}%
  \BibitemOpen
  \bibfield  {author} {\bibinfo {author} {\bibfnamefont {E.}~\bibnamefont
  {Knill}}, \bibinfo {author} {\bibfnamefont {R.}~\bibnamefont {Laflamme}}, \
  and\ \bibinfo {author} {\bibfnamefont {G.~J.}\ \bibnamefont {Milburn}},\
  }\href@noop {} {\bibfield  {journal} {\bibinfo  {journal} {Nature}\ }\textbf
  {\bibinfo {volume} {409}},\ \bibinfo {pages} {46} (\bibinfo {year}
  {2001})}\BibitemShut {NoStop}%
\bibitem [{\citenamefont {Ralph}\ \emph {et~al.}(2003)\citenamefont {Ralph},
  \citenamefont {Gilchrist}, \citenamefont {Milburn}, \citenamefont {Munro},\
  and\ \citenamefont {Glancy}}]{Ralph:2003jk}%
  \BibitemOpen
  \bibfield  {author} {\bibinfo {author} {\bibfnamefont {T.~C.}\ \bibnamefont
  {Ralph}}, \bibinfo {author} {\bibfnamefont {A.}~\bibnamefont {Gilchrist}},
  \bibinfo {author} {\bibfnamefont {G.~J.}\ \bibnamefont {Milburn}}, \bibinfo
  {author} {\bibfnamefont {W.~J.}\ \bibnamefont {Munro}}, \ and\ \bibinfo
  {author} {\bibfnamefont {S.}~\bibnamefont {Glancy}},\ }\href@noop {}
  {\bibfield  {journal} {\bibinfo  {journal} {Physical Review A}\ }\textbf
  {\bibinfo {volume} {68}},\ \bibinfo {pages} {042319} (\bibinfo {year}
  {2003})}\BibitemShut {NoStop}%
\bibitem [{\citenamefont {Lund}\ \emph {et~al.}(2008)\citenamefont {Lund},
  \citenamefont {Ralph},\ and\ \citenamefont {Haselgrove}}]{Anonymous:hr}%
  \BibitemOpen
  \bibfield  {author} {\bibinfo {author} {\bibfnamefont {A.~P.}\ \bibnamefont
  {Lund}}, \bibinfo {author} {\bibfnamefont {T.~C.}\ \bibnamefont {Ralph}}, \
  and\ \bibinfo {author} {\bibfnamefont {H.~L.}\ \bibnamefont {Haselgrove}},\
  }\href@noop {} {\bibfield  {journal} {\bibinfo  {journal} {Physical Review
  Letters}\ }\textbf {\bibinfo {volume} {100}},\ \bibinfo {pages} {030503}
  (\bibinfo {year} {2008})}\BibitemShut {NoStop}%
\bibitem [{\citenamefont {Leghtas}\ \emph {et~al.}(2013)\citenamefont
  {Leghtas}, \citenamefont {Kirchmair}, \citenamefont {Vlastakis},
  \citenamefont {Schoelkopf}, \citenamefont {Devoret},\ and\ \citenamefont
  {Mirrahimi}}]{Leghtas:2013ff}%
  \BibitemOpen
  \bibfield  {author} {\bibinfo {author} {\bibfnamefont {Z.}~\bibnamefont
  {Leghtas}}, \bibinfo {author} {\bibfnamefont {G.}~\bibnamefont {Kirchmair}},
  \bibinfo {author} {\bibfnamefont {B.}~\bibnamefont {Vlastakis}}, \bibinfo
  {author} {\bibfnamefont {R.~J.}\ \bibnamefont {Schoelkopf}}, \bibinfo
  {author} {\bibfnamefont {M.~H.}\ \bibnamefont {Devoret}}, \ and\ \bibinfo
  {author} {\bibfnamefont {M.}~\bibnamefont {Mirrahimi}},\ }\href@noop {}
  {\bibfield  {journal} {\bibinfo  {journal} {Physical Review Letters}\
  }\textbf {\bibinfo {volume} {111}},\ \bibinfo {pages} {120501} (\bibinfo
  {year} {2013})}\BibitemShut {NoStop}%
\bibitem [{\citenamefont {Mirrahimi}\ \emph {et~al.}(2014)\citenamefont
  {Mirrahimi}, \citenamefont {Leghtas}, \citenamefont {Albert}, \citenamefont
  {Touzard}, \citenamefont {Schoelkopf}, \citenamefont {Jiang},\ and\
  \citenamefont {Devoret}}]{Mirrahimi:2014js}%
  \BibitemOpen
  \bibfield  {author} {\bibinfo {author} {\bibfnamefont {M.}~\bibnamefont
  {Mirrahimi}}, \bibinfo {author} {\bibfnamefont {Z.}~\bibnamefont {Leghtas}},
  \bibinfo {author} {\bibfnamefont {V.~V.}\ \bibnamefont {Albert}}, \bibinfo
  {author} {\bibfnamefont {S.}~\bibnamefont {Touzard}}, \bibinfo {author}
  {\bibfnamefont {R.~J.}\ \bibnamefont {Schoelkopf}}, \bibinfo {author}
  {\bibfnamefont {L.}~\bibnamefont {Jiang}}, \ and\ \bibinfo {author}
  {\bibfnamefont {M.~H.}\ \bibnamefont {Devoret}},\ }\href@noop {} {\bibfield
  {journal} {\bibinfo  {journal} {New Journal of Physics}\ }\textbf {\bibinfo
  {volume} {16}},\ \bibinfo {pages} {045014} (\bibinfo {year}
  {2014})}\BibitemShut {NoStop}%
\bibitem [{\citenamefont {Gottesman}\ \emph {et~al.}(2001)\citenamefont
  {Gottesman}, \citenamefont {Kitaev},\ and\ \citenamefont
  {Preskill}}]{Gottesman:2001vk}%
  \BibitemOpen
  \bibfield  {author} {\bibinfo {author} {\bibfnamefont {D.}~\bibnamefont
  {Gottesman}}, \bibinfo {author} {\bibfnamefont {A.}~\bibnamefont {Kitaev}}, \
  and\ \bibinfo {author} {\bibfnamefont {J.}~\bibnamefont {Preskill}},\
  }\href@noop {} {\bibfield  {journal} {\bibinfo  {journal} {Physical Review
  A}\ }\textbf {\bibinfo {volume} {64}},\ \bibinfo {pages} {012310} (\bibinfo
  {year} {2001})}\BibitemShut {NoStop}%
\bibitem [{\citenamefont {Menicucci}(2014)}]{Menicucci:2014cx}%
  \BibitemOpen
  \bibfield  {author} {\bibinfo {author} {\bibfnamefont {N.~C.}\ \bibnamefont
  {Menicucci}},\ }\href@noop {} {\bibfield  {journal} {\bibinfo  {journal}
  {Physical Review Letters}\ }\textbf {\bibinfo {volume} {112}},\ \bibinfo
  {pages} {120504} (\bibinfo {year} {2014})}\BibitemShut {NoStop}%
\bibitem [{\citenamefont {Chuang}\ \emph {et~al.}(1997)\citenamefont {Chuang},
  \citenamefont {Leung},\ and\ \citenamefont {Yamamoto}}]{Chuang:1997dx}%
  \BibitemOpen
  \bibfield  {author} {\bibinfo {author} {\bibfnamefont {I.~L.}\ \bibnamefont
  {Chuang}}, \bibinfo {author} {\bibfnamefont {D.~W.}\ \bibnamefont {Leung}}, \
  and\ \bibinfo {author} {\bibfnamefont {Y.}~\bibnamefont {Yamamoto}},\
  }\href@noop {} {\bibfield  {journal} {\bibinfo  {journal} {Physical Review
  A}\ }\textbf {\bibinfo {volume} {56}},\ \bibinfo {pages} {1114} (\bibinfo
  {year} {1997})}\BibitemShut {NoStop}%
\bibitem [{\citenamefont {Ketterer}\ \emph {et~al.}(2015)\citenamefont
  {Ketterer}, \citenamefont {Keller}, \citenamefont {Walborn}, \citenamefont
  {Coudreau},\ and\ \citenamefont {Milman}}]{Ketterer:2015Code}%
  \BibitemOpen
  \bibfield  {author} {\bibinfo {author} {\bibfnamefont {A.}~\bibnamefont
  {Ketterer}}, \bibinfo {author} {\bibfnamefont {A.}~\bibnamefont {Keller}},
  \bibinfo {author} {\bibfnamefont {S.~P.}\ \bibnamefont {Walborn}}, \bibinfo
  {author} {\bibfnamefont {T.}~\bibnamefont {Coudreau}}, \ and\ \bibinfo
  {author} {\bibfnamefont {P.}~\bibnamefont {Milman}},\ }\href@noop {}
  {\bibfield  {journal} {\bibinfo  {journal} {preprint
  quant-ph/arXiv:1512.02957v1}\ } (\bibinfo {year} {2015})}\BibitemShut
  {NoStop}%
\bibitem [{\citenamefont {Michael}\ \emph {et~al.}(2016)\citenamefont
  {Michael}, \citenamefont {Silveri}, \citenamefont {Brierley}, \citenamefont
  {Albert}, \citenamefont {Salmilehto}, \citenamefont {Jiang},\ and\
  \citenamefont {Girvin}}]{Michael:2016code}%
  \BibitemOpen
  \bibfield  {author} {\bibinfo {author} {\bibfnamefont {M.~H.}\ \bibnamefont
  {Michael}}, \bibinfo {author} {\bibfnamefont {M.}~\bibnamefont {Silveri}},
  \bibinfo {author} {\bibfnamefont {R.~T.}\ \bibnamefont {Brierley}}, \bibinfo
  {author} {\bibfnamefont {V.~V.}\ \bibnamefont {Albert}}, \bibinfo {author}
  {\bibfnamefont {J.}~\bibnamefont {Salmilehto}}, \bibinfo {author}
  {\bibfnamefont {L.}~\bibnamefont {Jiang}}, \ and\ \bibinfo {author}
  {\bibfnamefont {S.~M.}\ \bibnamefont {Girvin}},\ }\href@noop {} {\bibfield
  {journal} {\bibinfo  {journal} {preprint quant-ph/arXiv1602.00008v1}\ }
  (\bibinfo {year} {2016})}\BibitemShut {NoStop}%
\bibitem [{\citenamefont {O{\textquoteright}Connell}\ \emph
  {et~al.}(2010)\citenamefont {O{\textquoteright}Connell}, \citenamefont
  {Hofheinz}, \citenamefont {Ansmann}, \citenamefont {Bialczak}, \citenamefont
  {Lenander}, \citenamefont {Lucero}, \citenamefont {Neeley}, \citenamefont
  {Sank}, \citenamefont {Wang}, \citenamefont {Weides}, \citenamefont {Wenner},
  \citenamefont {Martinis},\ and\ \citenamefont {Cleland}}]{OConnell:2010br}%
  \BibitemOpen
  \bibfield  {author} {\bibinfo {author} {\bibfnamefont {A.~D.}\ \bibnamefont
  {O{\textquoteright}Connell}}, \bibinfo {author} {\bibfnamefont
  {M.}~\bibnamefont {Hofheinz}}, \bibinfo {author} {\bibfnamefont
  {M.}~\bibnamefont {Ansmann}}, \bibinfo {author} {\bibfnamefont {R.~C.}\
  \bibnamefont {Bialczak}}, \bibinfo {author} {\bibfnamefont {M.}~\bibnamefont
  {Lenander}}, \bibinfo {author} {\bibfnamefont {E.}~\bibnamefont {Lucero}},
  \bibinfo {author} {\bibfnamefont {M.}~\bibnamefont {Neeley}}, \bibinfo
  {author} {\bibfnamefont {D.}~\bibnamefont {Sank}}, \bibinfo {author}
  {\bibfnamefont {H.}~\bibnamefont {Wang}}, \bibinfo {author} {\bibfnamefont
  {M.}~\bibnamefont {Weides}}, \bibinfo {author} {\bibfnamefont
  {J.}~\bibnamefont {Wenner}}, \bibinfo {author} {\bibfnamefont {J.~M.}\
  \bibnamefont {Martinis}}, \ and\ \bibinfo {author} {\bibfnamefont {A.~N.}\
  \bibnamefont {Cleland}},\ }\href@noop {} {\bibfield  {journal} {\bibinfo
  {journal} {Nature}\ }\textbf {\bibinfo {volume} {464}},\ \bibinfo {pages}
  {697} (\bibinfo {year} {2010})}\BibitemShut {NoStop}%
\bibitem [{\citenamefont {Cirac}\ \emph {et~al.}(1992)\citenamefont {Cirac},
  \citenamefont {Blatt}, \citenamefont {Zoller},\ and\ \citenamefont
  {Phillips}}]{Cirac:1992tp}%
  \BibitemOpen
  \bibfield  {author} {\bibinfo {author} {\bibfnamefont {J.}~\bibnamefont
  {Cirac}}, \bibinfo {author} {\bibfnamefont {R.}~\bibnamefont {Blatt}},
  \bibinfo {author} {\bibfnamefont {P.}~\bibnamefont {Zoller}}, \ and\ \bibinfo
  {author} {\bibfnamefont {W.~D.}\ \bibnamefont {Phillips}},\ }\href@noop {}
  {\bibfield  {journal} {\bibinfo  {journal} {Physical Review A}\ }\textbf
  {\bibinfo {volume} {46}},\ \bibinfo {pages} {2668} (\bibinfo {year}
  {1992})}\BibitemShut {NoStop}%
\bibitem [{\citenamefont {Marzoli}\ \emph {et~al.}(1994)\citenamefont
  {Marzoli}, \citenamefont {Cirac}, \citenamefont {Blatt},\ and\ \citenamefont
  {Zoller}}]{Marzoli:1994tz}%
  \BibitemOpen
  \bibfield  {author} {\bibinfo {author} {\bibfnamefont {I.}~\bibnamefont
  {Marzoli}}, \bibinfo {author} {\bibfnamefont {J.~I.}\ \bibnamefont {Cirac}},
  \bibinfo {author} {\bibfnamefont {R.}~\bibnamefont {Blatt}}, \ and\ \bibinfo
  {author} {\bibfnamefont {P.}~\bibnamefont {Zoller}},\ }\href@noop {}
  {\bibfield  {journal} {\bibinfo  {journal} {Physical Review A}\ }\textbf
  {\bibinfo {volume} {49}},\ \bibinfo {pages} {2771} (\bibinfo {year}
  {1994})}\BibitemShut {NoStop}%
\bibitem [{\citenamefont {Hopkins}\ \emph {et~al.}(2003)\citenamefont
  {Hopkins}, \citenamefont {Jacobs}, \citenamefont {Habib},\ and\ \citenamefont
  {Schwab}}]{Hopkins:2003hy}%
  \BibitemOpen
  \bibfield  {author} {\bibinfo {author} {\bibfnamefont {A.}~\bibnamefont
  {Hopkins}}, \bibinfo {author} {\bibfnamefont {K.}~\bibnamefont {Jacobs}},
  \bibinfo {author} {\bibfnamefont {S.}~\bibnamefont {Habib}}, \ and\ \bibinfo
  {author} {\bibfnamefont {K.}~\bibnamefont {Schwab}},\ }\href@noop {}
  {\bibfield  {journal} {\bibinfo  {journal} {Physical Review B}\ }\textbf
  {\bibinfo {volume} {68}},\ \bibinfo {pages} {235328} (\bibinfo {year}
  {2003})}\BibitemShut {NoStop}%
\bibitem [{\citenamefont {Wilson-Rae}\ \emph {et~al.}(2007)\citenamefont
  {Wilson-Rae}, \citenamefont {Nooshi}, \citenamefont {Zwerger},\ and\
  \citenamefont {Kippenberg}}]{WilsonRae:2007jp}%
  \BibitemOpen
  \bibfield  {author} {\bibinfo {author} {\bibfnamefont {I.}~\bibnamefont
  {Wilson-Rae}}, \bibinfo {author} {\bibfnamefont {N.}~\bibnamefont {Nooshi}},
  \bibinfo {author} {\bibfnamefont {W.}~\bibnamefont {Zwerger}}, \ and\
  \bibinfo {author} {\bibfnamefont {T.~J.}\ \bibnamefont {Kippenberg}},\
  }\href@noop {} {\bibfield  {journal} {\bibinfo  {journal} {Physical Review
  Letters}\ }\textbf {\bibinfo {volume} {99}},\ \bibinfo {pages} {093901}
  (\bibinfo {year} {2007})}\BibitemShut {NoStop}%
\bibitem [{\citenamefont {Marquardt}\ \emph {et~al.}(2007)\citenamefont
  {Marquardt}, \citenamefont {Chen}, \citenamefont {Clerk},\ and\ \citenamefont
  {Girvin}}]{Marquardt:2007dn}%
  \BibitemOpen
  \bibfield  {author} {\bibinfo {author} {\bibfnamefont {F.}~\bibnamefont
  {Marquardt}}, \bibinfo {author} {\bibfnamefont {J.~P.}\ \bibnamefont {Chen}},
  \bibinfo {author} {\bibfnamefont {A.~A.}\ \bibnamefont {Clerk}}, \ and\
  \bibinfo {author} {\bibfnamefont {S.~M.}\ \bibnamefont {Girvin}},\
  }\href@noop {} {\bibfield  {journal} {\bibinfo  {journal} {Physical Review
  Letters}\ }\textbf {\bibinfo {volume} {99}},\ \bibinfo {pages} {093902}
  (\bibinfo {year} {2007})}\BibitemShut {NoStop}%
\bibitem [{\citenamefont {Jaehne}\ \emph {et~al.}(2008)\citenamefont {Jaehne},
  \citenamefont {Hammerer},\ and\ \citenamefont {Wallquist}}]{Jaehne:2008gw}%
  \BibitemOpen
  \bibfield  {author} {\bibinfo {author} {\bibfnamefont {K.}~\bibnamefont
  {Jaehne}}, \bibinfo {author} {\bibfnamefont {K.}~\bibnamefont {Hammerer}}, \
  and\ \bibinfo {author} {\bibfnamefont {M.}~\bibnamefont {Wallquist}},\
  }\href@noop {} {\bibfield  {journal} {\bibinfo  {journal} {New Journal of
  Physics}\ }\textbf {\bibinfo {volume} {10}},\ \bibinfo {pages} {095019}
  (\bibinfo {year} {2008})}\BibitemShut {NoStop}%
\bibitem [{\citenamefont {Teufel}\ \emph {et~al.}(2011)\citenamefont {Teufel},
  \citenamefont {Donner}, \citenamefont {Li}, \citenamefont {Harlow},
  \citenamefont {Allman}, \citenamefont {Cicak}, \citenamefont {Sirois},
  \citenamefont {Whittaker}, \citenamefont {Lehnert},\ and\ \citenamefont
  {Simmonds}}]{Teufel:2011jg}%
  \BibitemOpen
  \bibfield  {author} {\bibinfo {author} {\bibfnamefont {J.~D.}\ \bibnamefont
  {Teufel}}, \bibinfo {author} {\bibfnamefont {T.}~\bibnamefont {Donner}},
  \bibinfo {author} {\bibfnamefont {D.}~\bibnamefont {Li}}, \bibinfo {author}
  {\bibfnamefont {J.~W.}\ \bibnamefont {Harlow}}, \bibinfo {author}
  {\bibfnamefont {M.~S.}\ \bibnamefont {Allman}}, \bibinfo {author}
  {\bibfnamefont {K.}~\bibnamefont {Cicak}}, \bibinfo {author} {\bibfnamefont
  {A.~J.}\ \bibnamefont {Sirois}}, \bibinfo {author} {\bibfnamefont {J.~D.}\
  \bibnamefont {Whittaker}}, \bibinfo {author} {\bibfnamefont {K.~W.}\
  \bibnamefont {Lehnert}}, \ and\ \bibinfo {author} {\bibfnamefont {R.~W.}\
  \bibnamefont {Simmonds}},\ }\href@noop {} {\bibfield  {journal} {\bibinfo
  {journal} {Nature}\ }\textbf {\bibinfo {volume} {475}},\ \bibinfo {pages}
  {359} (\bibinfo {year} {2011})}\BibitemShut {NoStop}%
\bibitem [{\citenamefont {Chan}\ \emph {et~al.}(2011)\citenamefont {Chan},
  \citenamefont {Alegre}, \citenamefont {Safavi-Naeini}, \citenamefont {Hill},
  \citenamefont {Krause}, \citenamefont {Gr{\"o}blacher}, \citenamefont
  {Aspelmeyer},\ and\ \citenamefont {Painter}}]{Chan:2011dy}%
  \BibitemOpen
  \bibfield  {author} {\bibinfo {author} {\bibfnamefont {J.}~\bibnamefont
  {Chan}}, \bibinfo {author} {\bibfnamefont {T.~P.~M.}\ \bibnamefont {Alegre}},
  \bibinfo {author} {\bibfnamefont {A.~H.}\ \bibnamefont {Safavi-Naeini}},
  \bibinfo {author} {\bibfnamefont {J.~T.}\ \bibnamefont {Hill}}, \bibinfo
  {author} {\bibfnamefont {A.}~\bibnamefont {Krause}}, \bibinfo {author}
  {\bibfnamefont {S.}~\bibnamefont {Gr{\"o}blacher}}, \bibinfo {author}
  {\bibfnamefont {M.}~\bibnamefont {Aspelmeyer}}, \ and\ \bibinfo {author}
  {\bibfnamefont {O.}~\bibnamefont {Painter}},\ }\href@noop {} {\bibfield
  {journal} {\bibinfo  {journal} {Nature}\ }\textbf {\bibinfo {volume} {478}},\
  \bibinfo {pages} {89} (\bibinfo {year} {2011})}\BibitemShut {NoStop}%
\bibitem [{\citenamefont {Machnes}\ \emph {et~al.}(2010)\citenamefont
  {Machnes}, \citenamefont {Plenio}, \citenamefont {Reznik}, \citenamefont
  {Steane},\ and\ \citenamefont {Retzker}}]{Machnes:2010bg}%
  \BibitemOpen
  \bibfield  {author} {\bibinfo {author} {\bibfnamefont {S.}~\bibnamefont
  {Machnes}}, \bibinfo {author} {\bibfnamefont {M.~B.}\ \bibnamefont {Plenio}},
  \bibinfo {author} {\bibfnamefont {B.}~\bibnamefont {Reznik}}, \bibinfo
  {author} {\bibfnamefont {A.~M.}\ \bibnamefont {Steane}}, \ and\ \bibinfo
  {author} {\bibfnamefont {A.}~\bibnamefont {Retzker}},\ }\href@noop {}
  {\bibfield  {journal} {\bibinfo  {journal} {Physical Review Letters}\
  }\textbf {\bibinfo {volume} {104}},\ \bibinfo {pages} {183001} (\bibinfo
  {year} {2010})}\BibitemShut {NoStop}%
\bibitem [{\citenamefont {Machnes}\ \emph {et~al.}(2012)\citenamefont
  {Machnes}, \citenamefont {Cerrillo}, \citenamefont {Aspelmeyer},
  \citenamefont {Wieczorek}, \citenamefont {Plenio},\ and\ \citenamefont
  {Retzker}}]{Machnes:2012bg}%
  \BibitemOpen
  \bibfield  {author} {\bibinfo {author} {\bibfnamefont {S.}~\bibnamefont
  {Machnes}}, \bibinfo {author} {\bibfnamefont {J.}~\bibnamefont {Cerrillo}},
  \bibinfo {author} {\bibfnamefont {M.}~\bibnamefont {Aspelmeyer}}, \bibinfo
  {author} {\bibfnamefont {W.}~\bibnamefont {Wieczorek}}, \bibinfo {author}
  {\bibfnamefont {M.~B.}\ \bibnamefont {Plenio}}, \ and\ \bibinfo {author}
  {\bibfnamefont {A.}~\bibnamefont {Retzker}},\ }\href@noop {} {\bibfield
  {journal} {\bibinfo  {journal} {Physical Review Letters}\ }\textbf {\bibinfo
  {volume} {108}},\ \bibinfo {pages} {153601} (\bibinfo {year}
  {2012})}\BibitemShut {NoStop}%
\bibitem [{\citenamefont {Martin}\ \emph {et~al.}(2004)\citenamefont {Martin},
  \citenamefont {Shnirman}, \citenamefont {Tian},\ and\ \citenamefont
  {Zoller}}]{Martin:2004kv}%
  \BibitemOpen
  \bibfield  {author} {\bibinfo {author} {\bibfnamefont {I.}~\bibnamefont
  {Martin}}, \bibinfo {author} {\bibfnamefont {A.}~\bibnamefont {Shnirman}},
  \bibinfo {author} {\bibfnamefont {L.}~\bibnamefont {Tian}}, \ and\ \bibinfo
  {author} {\bibfnamefont {P.}~\bibnamefont {Zoller}},\ }\href@noop {}
  {\bibfield  {journal} {\bibinfo  {journal} {Physical Review B}\ }\textbf
  {\bibinfo {volume} {69}},\ \bibinfo {pages} {125339} (\bibinfo {year}
  {2004})}\BibitemShut {NoStop}%
\bibitem [{\citenamefont {Wilson-Rae}\ \emph {et~al.}(2004)\citenamefont
  {Wilson-Rae}, \citenamefont {Zoller},\ and\ \citenamefont
  {Imamo{\u{g}}lu}}]{WilsonRae:2004il}%
  \BibitemOpen
  \bibfield  {author} {\bibinfo {author} {\bibfnamefont {I.}~\bibnamefont
  {Wilson-Rae}}, \bibinfo {author} {\bibfnamefont {P.}~\bibnamefont {Zoller}},
  \ and\ \bibinfo {author} {\bibfnamefont {A.}~\bibnamefont {Imamo{\u{g}}lu}},\
  }\href@noop {} {\bibfield  {journal} {\bibinfo  {journal} {Physical Review
  Letters}\ }\textbf {\bibinfo {volume} {92}},\ \bibinfo {pages} {075507}
  (\bibinfo {year} {2004})}\BibitemShut {NoStop}%
\bibitem [{\citenamefont {Xia}\ and\ \citenamefont {Evers}(2009)}]{Xia:2009kl}%
  \BibitemOpen
  \bibfield  {author} {\bibinfo {author} {\bibfnamefont {K.}~\bibnamefont
  {Xia}}\ and\ \bibinfo {author} {\bibfnamefont {J.}~\bibnamefont {Evers}},\
  }\href@noop {} {\bibfield  {journal} {\bibinfo  {journal} {Physical Review
  Letters}\ }\textbf {\bibinfo {volume} {103}},\ \bibinfo {pages} {227203}
  (\bibinfo {year} {2009})}\BibitemShut {NoStop}%
\bibitem [{\citenamefont {Rabl}(2010)}]{Rabl:2010cm}%
  \BibitemOpen
  \bibfield  {author} {\bibinfo {author} {\bibfnamefont {P.}~\bibnamefont
  {Rabl}},\ }\href@noop {} {\bibfield  {journal} {\bibinfo  {journal} {Physical
  Review B}\ }\textbf {\bibinfo {volume} {82}},\ \bibinfo {pages} {165320}
  (\bibinfo {year} {2010})}\BibitemShut {NoStop}%
\bibitem [{\citenamefont {Kepesidis}\ \emph {et~al.}(2013)\citenamefont
  {Kepesidis}, \citenamefont {Bennett}, \citenamefont {Portolan}, \citenamefont
  {Lukin},\ and\ \citenamefont {Rabl}}]{Kepesidis:2013fv}%
  \BibitemOpen
  \bibfield  {author} {\bibinfo {author} {\bibfnamefont {K.~V.}\ \bibnamefont
  {Kepesidis}}, \bibinfo {author} {\bibfnamefont {S.~D.}\ \bibnamefont
  {Bennett}}, \bibinfo {author} {\bibfnamefont {S.}~\bibnamefont {Portolan}},
  \bibinfo {author} {\bibfnamefont {M.~D.}\ \bibnamefont {Lukin}}, \ and\
  \bibinfo {author} {\bibfnamefont {P.}~\bibnamefont {Rabl}},\ }\href@noop {}
  {\bibfield  {journal} {\bibinfo  {journal} {Physical Review B}\ }\textbf
  {\bibinfo {volume} {88}},\ \bibinfo {pages} {064105} (\bibinfo {year}
  {2013})}\BibitemShut {NoStop}%
\bibitem [{\citenamefont {Lau}\ and\ \citenamefont
  {Plenio}(2016{\natexlab{a}})}]{Lau:2016cool}%
  \BibitemOpen
  \bibfield  {author} {\bibinfo {author} {\bibfnamefont {H.-K.}\ \bibnamefont
  {Lau}}\ and\ \bibinfo {author} {\bibfnamefont {M.~B.}\ \bibnamefont
  {Plenio}},\ }\href@noop {} {\bibfield  {journal} {\bibinfo  {journal}
  {Physical Review B}\ }\textbf {\bibinfo {volume} {94}},\ \bibinfo {pages}
  {054305} (\bibinfo {year} {2016}{\natexlab{a}})}\BibitemShut {NoStop}%
\bibitem [{\citenamefont {Jeong}\ and\ \citenamefont
  {Ralph}(2006)}]{Jeong:2006gs}%
  \BibitemOpen
  \bibfield  {author} {\bibinfo {author} {\bibfnamefont {H.}~\bibnamefont
  {Jeong}}\ and\ \bibinfo {author} {\bibfnamefont {T.~C.}\ \bibnamefont
  {Ralph}},\ }\href@noop {} {\bibfield  {journal} {\bibinfo  {journal}
  {Physical Review Letters}\ }\textbf {\bibinfo {volume} {97}},\ \bibinfo
  {pages} {100401} (\bibinfo {year} {2006})}\BibitemShut {NoStop}%
\bibitem [{\citenamefont {Jeong}\ and\ \citenamefont
  {Ralph}(2007)}]{Jeong:2007bc}%
  \BibitemOpen
  \bibfield  {author} {\bibinfo {author} {\bibfnamefont {H.}~\bibnamefont
  {Jeong}}\ and\ \bibinfo {author} {\bibfnamefont {T.~C.}\ \bibnamefont
  {Ralph}},\ }\href@noop {} {\bibfield  {journal} {\bibinfo  {journal}
  {Physical Review A}\ }\textbf {\bibinfo {volume} {76}},\ \bibinfo {pages}
  {042103} (\bibinfo {year} {2007})}\BibitemShut {NoStop}%
\bibitem [{\citenamefont {Grace}\ \emph {et~al.}(2006)\citenamefont {Grace},
  \citenamefont {Brif}, \citenamefont {Rabitz}, \citenamefont {Walmsley},
  \citenamefont {Kosut},\ and\ \citenamefont {Lidar}}]{Grace:2006gc}%
  \BibitemOpen
  \bibfield  {author} {\bibinfo {author} {\bibfnamefont {M.}~\bibnamefont
  {Grace}}, \bibinfo {author} {\bibfnamefont {C.}~\bibnamefont {Brif}},
  \bibinfo {author} {\bibfnamefont {H.}~\bibnamefont {Rabitz}}, \bibinfo
  {author} {\bibfnamefont {I.}~\bibnamefont {Walmsley}}, \bibinfo {author}
  {\bibfnamefont {R.}~\bibnamefont {Kosut}}, \ and\ \bibinfo {author}
  {\bibfnamefont {D.}~\bibnamefont {Lidar}},\ }\href@noop {} {\bibfield
  {journal} {\bibinfo  {journal} {New Journal of Physics}\ }\textbf {\bibinfo
  {volume} {8}},\ \bibinfo {pages} {35} (\bibinfo {year} {2006})}\BibitemShut
  {NoStop}%
\bibitem [{\citenamefont {Lau}\ and\ \citenamefont
  {Plenio}(2016{\natexlab{b}})}]{Lau:2016UUQC}%
  \BibitemOpen
  \bibfield  {author} {\bibinfo {author} {\bibfnamefont {H.-K.}\ \bibnamefont
  {Lau}}\ and\ \bibinfo {author} {\bibfnamefont {M.~B.}\ \bibnamefont
  {Plenio}},\ }\href@noop {} {\bibfield  {journal} {\bibinfo  {journal}
  {preprint quant-ph/arXiv:1605.09278}\ } (\bibinfo {year}
  {2016}{\natexlab{b}})}\BibitemShut {NoStop}%
\bibitem [{\citenamefont {Gershenfeld}\ and\ \citenamefont
  {Chuang}(1997)}]{Gershenfeld:1997ie}%
  \BibitemOpen
  \bibfield  {author} {\bibinfo {author} {\bibfnamefont {N.~A.}\ \bibnamefont
  {Gershenfeld}}\ and\ \bibinfo {author} {\bibfnamefont {I.~L.}\ \bibnamefont
  {Chuang}},\ }\href@noop {} {\bibfield  {journal} {\bibinfo  {journal}
  {Science}\ }\textbf {\bibinfo {volume} {275}},\ \bibinfo {pages} {350}
  (\bibinfo {year} {1997})}\BibitemShut {NoStop}%
\bibitem [{\citenamefont {Knill}\ and\ \citenamefont
  {Laflamme}(1998)}]{Knill:1998us}%
  \BibitemOpen
  \bibfield  {author} {\bibinfo {author} {\bibfnamefont {E.}~\bibnamefont
  {Knill}}\ and\ \bibinfo {author} {\bibfnamefont {R.}~\bibnamefont
  {Laflamme}},\ }\href@noop {} {\bibfield  {journal} {\bibinfo  {journal}
  {Physical Review Letters}\ }\textbf {\bibinfo {volume} {81}},\ \bibinfo
  {pages} {5672} (\bibinfo {year} {1998})}\BibitemShut {NoStop}%
\bibitem [{\citenamefont {Liu}\ \emph {et~al.}(2016)\citenamefont {Liu},
  \citenamefont {Thompson}, \citenamefont {Weedbrook}, \citenamefont {Lloyd},
  \citenamefont {Vedral}, \citenamefont {Gu},\ and\ \citenamefont
  {Modi}}]{PhysRevA.93.052304}%
  \BibitemOpen
  \bibfield  {author} {\bibinfo {author} {\bibfnamefont {N.}~\bibnamefont
  {Liu}}, \bibinfo {author} {\bibfnamefont {J.}~\bibnamefont {Thompson}},
  \bibinfo {author} {\bibfnamefont {C.}~\bibnamefont {Weedbrook}}, \bibinfo
  {author} {\bibfnamefont {S.}~\bibnamefont {Lloyd}}, \bibinfo {author}
  {\bibfnamefont {V.}~\bibnamefont {Vedral}}, \bibinfo {author} {\bibfnamefont
  {M.}~\bibnamefont {Gu}}, \ and\ \bibinfo {author} {\bibfnamefont
  {K.}~\bibnamefont {Modi}},\ }\href@noop {} {\bibfield  {journal} {\bibinfo
  {journal} {Phys. Rev. A}\ }\textbf {\bibinfo {volume} {93}},\ \bibinfo
  {pages} {052304} (\bibinfo {year} {2016})}\BibitemShut {NoStop}%
\bibitem [{Note1()}]{Note1}%
  \BibitemOpen
  \bibinfo {note} {The $\protect \mathaccentV {hat}05E{Y}_L$ operator is
  defined to be $i\protect \mathaccentV {hat}05E{X}_L \protect \mathaccentV
  {hat}05E{Z}_L$ for any $\protect \mathaccentV {hat}05E{X}_L$ and $\protect
  \mathaccentV {hat}05E{Z}_L$.}\BibitemShut {Stop}%
\bibitem [{\citenamefont {Lau}\ \emph {et~al.}(2016)\citenamefont {Lau},
  \citenamefont {Pooser}, \citenamefont {Siopsis},\ and\ \citenamefont
  {Weedbrook}}]{Lau:2016QML}%
  \BibitemOpen
  \bibfield  {author} {\bibinfo {author} {\bibfnamefont {H.-K.}\ \bibnamefont
  {Lau}}, \bibinfo {author} {\bibfnamefont {R.}~\bibnamefont {Pooser}},
  \bibinfo {author} {\bibfnamefont {G.}~\bibnamefont {Siopsis}}, \ and\
  \bibinfo {author} {\bibfnamefont {C.}~\bibnamefont {Weedbrook}},\ }\href@noop
  {} {\bibfield  {journal} {\bibinfo  {journal} {preprint
  quant-ph/arXiv:1603.06222}\ } (\bibinfo {year} {2016})}\BibitemShut {NoStop}%
\bibitem [{\citenamefont {Marvian}\ and\ \citenamefont
  {Lloyd}(2016)}]{Iman:2016UQE}%
  \BibitemOpen
  \bibfield  {author} {\bibinfo {author} {\bibfnamefont {I.}~\bibnamefont
  {Marvian}}\ and\ \bibinfo {author} {\bibfnamefont {S.}~\bibnamefont
  {Lloyd}},\ }\href@noop {} {\bibfield  {journal} {\bibinfo  {journal}
  {preprint quant-ph/arXiv:1606.02734}\ } (\bibinfo {year} {2016})}\BibitemShut
  {NoStop}%
\bibitem [{\citenamefont {Plick}\ \emph {et~al.}(2010)\citenamefont {Plick},
  \citenamefont {Anisimov}, \citenamefont {Dowling}, \citenamefont {Lee},\ and\
  \citenamefont {Agarwal}}]{Plick:2010fa}%
  \BibitemOpen
  \bibfield  {author} {\bibinfo {author} {\bibfnamefont {W.~N.}\ \bibnamefont
  {Plick}}, \bibinfo {author} {\bibfnamefont {P.~M.}\ \bibnamefont {Anisimov}},
  \bibinfo {author} {\bibfnamefont {J.~P.}\ \bibnamefont {Dowling}}, \bibinfo
  {author} {\bibfnamefont {H.}~\bibnamefont {Lee}}, \ and\ \bibinfo {author}
  {\bibfnamefont {G.~S.}\ \bibnamefont {Agarwal}},\ }\href@noop {} {\bibfield
  {journal} {\bibinfo  {journal} {New Journal of Physics}\ }\textbf {\bibinfo
  {volume} {12}},\ \bibinfo {pages} {113025} (\bibinfo {year}
  {2010})}\BibitemShut {NoStop}%
\bibitem [{\citenamefont {Leonhardt}(2010)}]{book:Ulf}%
  \BibitemOpen
  \bibfield  {author} {\bibinfo {author} {\bibfnamefont {U.}~\bibnamefont
  {Leonhardt}},\ }\href@noop {} {\emph {\bibinfo {title} {Essential Quantum
  Optics, From Quantum Measurements to Black Holes}}}\ (\bibinfo  {publisher}
  {Cambridge University Press},\ \bibinfo {year} {2010})\BibitemShut {NoStop}%
\bibitem [{\citenamefont {Anders}\ \emph {et~al.}(2010)\citenamefont {Anders},
  \citenamefont {Oi}, \citenamefont {Kashefi}, \citenamefont {Browne},\ and\
  \citenamefont {Andersson}}]{Anders:2010ic}%
  \BibitemOpen
  \bibfield  {author} {\bibinfo {author} {\bibfnamefont {J.}~\bibnamefont
  {Anders}}, \bibinfo {author} {\bibfnamefont {D.~K.~L.}\ \bibnamefont {Oi}},
  \bibinfo {author} {\bibfnamefont {E.}~\bibnamefont {Kashefi}}, \bibinfo
  {author} {\bibfnamefont {D.~E.}\ \bibnamefont {Browne}}, \ and\ \bibinfo
  {author} {\bibfnamefont {E.}~\bibnamefont {Andersson}},\ }\href@noop {}
  {\bibfield  {journal} {\bibinfo  {journal} {Physical Review A}\ }\textbf
  {\bibinfo {volume} {82}},\ \bibinfo {pages} {020301} (\bibinfo {year}
  {2010})}\BibitemShut {NoStop}%
\bibitem [{\citenamefont {Blais}\ \emph {et~al.}(2004)\citenamefont {Blais},
  \citenamefont {Huang}, \citenamefont {Wallraff}, \citenamefont {Girvin},\
  and\ \citenamefont {Schoelkopf}}]{Blais:2004kn}%
  \BibitemOpen
  \bibfield  {author} {\bibinfo {author} {\bibfnamefont {A.}~\bibnamefont
  {Blais}}, \bibinfo {author} {\bibfnamefont {R.-S.}\ \bibnamefont {Huang}},
  \bibinfo {author} {\bibfnamefont {A.}~\bibnamefont {Wallraff}}, \bibinfo
  {author} {\bibfnamefont {S.~M.}\ \bibnamefont {Girvin}}, \ and\ \bibinfo
  {author} {\bibfnamefont {R.~J.}\ \bibnamefont {Schoelkopf}},\ }\href@noop {}
  {\bibfield  {journal} {\bibinfo  {journal} {Physical Review A}\ }\textbf
  {\bibinfo {volume} {69}},\ \bibinfo {pages} {062320} (\bibinfo {year}
  {2004})}\BibitemShut {NoStop}%
\bibitem [{\citenamefont {Rabl}\ \emph {et~al.}(2009)\citenamefont {Rabl},
  \citenamefont {Cappellaro}, \citenamefont {Dutt}, \citenamefont {Jiang},
  \citenamefont {Maze},\ and\ \citenamefont {Lukin}}]{Rabl:2009fz}%
  \BibitemOpen
  \bibfield  {author} {\bibinfo {author} {\bibfnamefont {P.}~\bibnamefont
  {Rabl}}, \bibinfo {author} {\bibfnamefont {P.}~\bibnamefont {Cappellaro}},
  \bibinfo {author} {\bibfnamefont {M.~V.~G.}\ \bibnamefont {Dutt}}, \bibinfo
  {author} {\bibfnamefont {L.}~\bibnamefont {Jiang}}, \bibinfo {author}
  {\bibfnamefont {J.~R.}\ \bibnamefont {Maze}}, \ and\ \bibinfo {author}
  {\bibfnamefont {M.~D.}\ \bibnamefont {Lukin}},\ }\href@noop {} {\bibfield
  {journal} {\bibinfo  {journal} {Physical Review B}\ }\textbf {\bibinfo
  {volume} {79}},\ \bibinfo {pages} {041302} (\bibinfo {year}
  {2009})}\BibitemShut {NoStop}%
\bibitem [{\citenamefont {Lloyd}(2003)}]{Lloyd:2003HQC}%
  \BibitemOpen
  \bibfield  {author} {\bibinfo {author} {\bibfnamefont {S.}~\bibnamefont
  {Lloyd}},\ }\enquote {\bibinfo {title} {Hybrid quantum computing},}\ in\
  \href {\doibase 10.1007/978-94-015-1258-9_5} {\emph {\bibinfo {booktitle}
  {Quantum Information with Continuous Variables}}},\ \bibinfo {editor} {edited
  by\ \bibinfo {editor} {\bibfnamefont {S.~L.}\ \bibnamefont {Braunstein}}\
  and\ \bibinfo {editor} {\bibfnamefont {A.~K.}\ \bibnamefont {Pati}}}\
  (\bibinfo  {publisher} {Springer Netherlands},\ \bibinfo {address}
  {Dordrecht},\ \bibinfo {year} {2003})\ pp.\ \bibinfo {pages}
  {37--45}\BibitemShut {NoStop}%
\bibitem [{\citenamefont {Landauer}(1961)}]{LANDAUER:1961wc}%
  \BibitemOpen
  \bibfield  {author} {\bibinfo {author} {\bibfnamefont {R.}~\bibnamefont
  {Landauer}},\ }\href@noop {} {\bibfield  {journal} {\bibinfo  {journal} {IBM
  Journal of Research and Development}\ }\textbf {\bibinfo {volume} {5}},\
  \bibinfo {pages} {183} (\bibinfo {year} {1961})}\BibitemShut {NoStop}%
\bibitem [{\citenamefont {Jacobs}(2012)}]{Jacobs:2012eh}%
  \BibitemOpen
  \bibfield  {author} {\bibinfo {author} {\bibfnamefont {K.}~\bibnamefont
  {Jacobs}},\ }\href@noop {} {\bibfield  {journal} {\bibinfo  {journal}
  {Physical Review E}\ }\textbf {\bibinfo {volume} {86}},\ \bibinfo {pages}
  {040106} (\bibinfo {year} {2012})}\BibitemShut {NoStop}%
\bibitem [{\citenamefont {Zanardi}\ \emph {et~al.}(2004)\citenamefont
  {Zanardi}, \citenamefont {Lidar},\ and\ \citenamefont
  {Lloyd}}]{Zanardi:2004kb}%
  \BibitemOpen
  \bibfield  {author} {\bibinfo {author} {\bibfnamefont {P.}~\bibnamefont
  {Zanardi}}, \bibinfo {author} {\bibfnamefont {D.~A.}\ \bibnamefont {Lidar}},
  \ and\ \bibinfo {author} {\bibfnamefont {S.}~\bibnamefont {Lloyd}},\
  }\href@noop {} {\bibfield  {journal} {\bibinfo  {journal} {Physical Review
  Letters}\ }\textbf {\bibinfo {volume} {92}},\ \bibinfo {pages} {060402}
  (\bibinfo {year} {2004})}\BibitemShut {NoStop}%
\bibitem [{\citenamefont {Knill}\ \emph {et~al.}(2000)\citenamefont {Knill},
  \citenamefont {Laflamme},\ and\ \citenamefont {Viola}}]{Knill:2000wt}%
  \BibitemOpen
  \bibfield  {author} {\bibinfo {author} {\bibfnamefont {E.}~\bibnamefont
  {Knill}}, \bibinfo {author} {\bibfnamefont {R.}~\bibnamefont {Laflamme}}, \
  and\ \bibinfo {author} {\bibfnamefont {L.}~\bibnamefont {Viola}},\
  }\href@noop {} {\bibfield  {journal} {\bibinfo  {journal} {Physical Review
  Letters}\ }\textbf {\bibinfo {volume} {84}},\ \bibinfo {pages} {2525}
  (\bibinfo {year} {2000})}\BibitemShut {NoStop}%
\bibitem [{\citenamefont {de~Filippo}(2000)}]{2000PhRvA..62e2307D}%
  \BibitemOpen
  \bibfield  {author} {\bibinfo {author} {\bibfnamefont {S.}~\bibnamefont
  {de~Filippo}},\ }\href@noop {} {\bibfield  {journal} {\bibinfo  {journal}
  {Physical Review A}\ }\textbf {\bibinfo {volume} {62}},\ \bibinfo {pages}
  {052307} (\bibinfo {year} {2000})}\BibitemShut {NoStop}%
\bibitem [{\citenamefont {Zanardi}(2000)}]{Zanardi:2001vo}%
  \BibitemOpen
  \bibfield  {author} {\bibinfo {author} {\bibfnamefont {P.}~\bibnamefont
  {Zanardi}},\ }\href@noop {} {\bibfield  {journal} {\bibinfo  {journal}
  {Physical Review A}\ }\textbf {\bibinfo {volume} {63}},\ \bibinfo {pages}
  {012301} (\bibinfo {year} {2000})}\BibitemShut {NoStop}%
\bibitem [{\citenamefont {Kempe}\ \emph {et~al.}(2001)\citenamefont {Kempe},
  \citenamefont {Bacon}, \citenamefont {Lidar},\ and\ \citenamefont
  {Whaley}}]{Kempe:2001bg}%
  \BibitemOpen
  \bibfield  {author} {\bibinfo {author} {\bibfnamefont {J.}~\bibnamefont
  {Kempe}}, \bibinfo {author} {\bibfnamefont {D.}~\bibnamefont {Bacon}},
  \bibinfo {author} {\bibfnamefont {D.~A.}\ \bibnamefont {Lidar}}, \ and\
  \bibinfo {author} {\bibfnamefont {K.~B.}\ \bibnamefont {Whaley}},\
  }\href@noop {} {\bibfield  {journal} {\bibinfo  {journal} {Physical Review
  A}\ }\textbf {\bibinfo {volume} {63}},\ \bibinfo {pages} {042307} (\bibinfo
  {year} {2001})}\BibitemShut {NoStop}%
\bibitem [{\citenamefont {Kribs}\ \emph {et~al.}(2005)\citenamefont {Kribs},
  \citenamefont {Laflamme},\ and\ \citenamefont {Poulin}}]{Kribs:2005id}%
  \BibitemOpen
  \bibfield  {author} {\bibinfo {author} {\bibfnamefont {D.}~\bibnamefont
  {Kribs}}, \bibinfo {author} {\bibfnamefont {R.}~\bibnamefont {Laflamme}}, \
  and\ \bibinfo {author} {\bibfnamefont {D.}~\bibnamefont {Poulin}},\
  }\href@noop {} {\bibfield  {journal} {\bibinfo  {journal} {Physical Review
  Letters}\ }\textbf {\bibinfo {volume} {94}},\ \bibinfo {pages} {180501}
  (\bibinfo {year} {2005})}\BibitemShut {NoStop}%
\bibitem [{\citenamefont {Lidar}\ \emph {et~al.}(1998)\citenamefont {Lidar},
  \citenamefont {Chuang},\ and\ \citenamefont {Whaley}}]{Lidar:1998wl}%
  \BibitemOpen
  \bibfield  {author} {\bibinfo {author} {\bibfnamefont {D.}~\bibnamefont
  {Lidar}}, \bibinfo {author} {\bibfnamefont {I.}~\bibnamefont {Chuang}}, \
  and\ \bibinfo {author} {\bibfnamefont {K.}~\bibnamefont {Whaley}},\
  }\href@noop {} {\bibfield  {journal} {\bibinfo  {journal} {Physical Review
  Letters}\ }\textbf {\bibinfo {volume} {81}},\ \bibinfo {pages} {2594}
  (\bibinfo {year} {1998})}\BibitemShut {NoStop}%
\bibitem [{\citenamefont {Dalibard}\ \emph {et~al.}(1992)\citenamefont
  {Dalibard}, \citenamefont {Castin},\ and\ \citenamefont
  {Molmer}}]{Dalibard:1992ub}%
  \BibitemOpen
  \bibfield  {author} {\bibinfo {author} {\bibfnamefont {J.}~\bibnamefont
  {Dalibard}}, \bibinfo {author} {\bibfnamefont {Y.}~\bibnamefont {Castin}}, \
  and\ \bibinfo {author} {\bibfnamefont {K.}~\bibnamefont {Molmer}},\
  }\href@noop {} {\bibfield  {journal} {\bibinfo  {journal} {Physical Review
  Letters}\ }\textbf {\bibinfo {volume} {68}},\ \bibinfo {pages} {580}
  (\bibinfo {year} {1992})}\BibitemShut {NoStop}%
\bibitem [{\citenamefont {Plenio}\ and\ \citenamefont
  {Knight}(1998)}]{Plenio:1998jk}%
  \BibitemOpen
  \bibfield  {author} {\bibinfo {author} {\bibfnamefont {M.~B.}\ \bibnamefont
  {Plenio}}\ and\ \bibinfo {author} {\bibfnamefont {P.~L.}\ \bibnamefont
  {Knight}},\ }\href@noop {} {\bibfield  {journal} {\bibinfo  {journal}
  {Reviews of modern physics}\ }\textbf {\bibinfo {volume} {70}},\ \bibinfo
  {pages} {101} (\bibinfo {year} {1998})}\BibitemShut {NoStop}%
\end{thebibliography}%

\appendix

\section{Appendix I: Mixed-State Quantum Computing}

Any unitary transformation $\mathcal{U}$ can be constructed from a series of computational steps, $\mathcal{U}=\mathcal{U}_1\mathcal{U}_2\ldots$, and each step can be decomposed into a series of universal logic gates in Eq.~(\ref{eq:U_gate}),
\begin{equation}
\mathcal{U}_q=\Big(\prod_{k=1}^K e^{i\phi^{[q]}_k \hat{Z}_{L_k}} \Big)\Big(\prod_{k=1}^Ke^{i\theta^{[q]}_k \hat{X}_{L_k}}\Big)\Big(\prod_{k=2}^K e^{i\gamma^{[q]}_k \hat{Z}_{L_k}\hat{Z}_{L_{k-1}}}\Big)~,
\end{equation}
where $\phi^{[q]}_k$, $\theta^{[q]}_k$, and $\gamma^{[q]}_k$ are the rotation angles of the $k$th qubit in the $q$th computational step.  

In MSUQC, each physical operation in Eq.~(\ref{eq:U_gate}) can be expressed as a sum of logical operations in each basis \cite{Grace:2006gc}.  For example, the mixed-state Pauli-Z rotation, $\hat{U}_Z(\phi^{[q]}_k)$, can be expressed as
\begin{eqnarray}
e^{i\phi^{[q]}_k \hat{Z}_L}&=& \cos\phi^{[q]}_k \hat{\mathbb{I}} + i\sin\phi^{[q]}_k \hat{Z}_L  \nonumber \\
&=& \sum_n \cos\phi^{[q]}_k \hat{\mathbb{I}}^{(n)} + i \sin\phi^{[q]}_k \hat{Z}_L^{(n)}~,
\end{eqnarray}
where $\hat{\mathbb{I}}^{(n)} = |\psi_0^{(n)}\rangle \langle\psi_0^{(n)}| + |\psi_1^{(n)}\rangle \langle\psi_1^{(n)}|$.  When applying on a single-qumode state in the  $n$th basis, we get
\begin{equation}
e^{i \phi^{[q]}_k \hat{Z}_L}|\psi^{(n)}\rangle = e^{i \phi^{[q]}_k \hat{Z}_L^{(n)}}|\psi^{(n)}\rangle~.
\end{equation}
This operation implements a Pauli-$Z$ rotation with angle $\phi^{[q]}_k$ in the $n$th basis, and the resultant state is a superposition of the basis states $|\psi_0^{(n)}\rangle$ and $|\psi_1^{(n)}\rangle$ only.  Similarly for $\hat{U}_X(\theta^{[q]}_k)$ and $\hat{U}_{ZZ}(\gamma^{[q]}_k)$, 
\begin{eqnarray}
e^{i \theta^{[q]}_k \hat{X}_L}|\psi^{(n)}\rangle &=& e^{i \theta^{[q]}_k \hat{X}_L^{(n)}}|\psi^{(n)}\rangle~, \\
e^{i \gamma^{[q]}_k \hat{Z}_L \otimes \hat{Z}_L}|\psi^{(n)}\rangle|\psi^{(n')}\rangle &=& e^{i \gamma^{[q]}_k  \hat{Z}_L^{(n)}\otimes\hat{Z}_L^{(n')}}|\psi^{(n)}\rangle|\psi^{(n')}\rangle~. \nonumber \\
\end{eqnarray}
The rotation angles are the same for each basis, and the resultant state remains in the original bases.

In the computational task, the initial state $\rho_0 = \sum_{\vec{n}} p_{\vec{n}} |\Psi_0^{\vec{n}}\rangle\langle\Psi_0^{\vec{n}}|$ can be viewed as an ensemble of pure states $ |\Psi_0^{\vec{n}}\rangle$, which each appears with probability $p_{\vec{n}}$.
Although $\vec{n}=\{n_1,\ldots,n_K \}$ is unknown in each computation, it has been implicitly defined by the initial state $|\Psi_0^{\vec{n}}\rangle$, and does not change during the logical operations.  If the projective measurement basis is the same, i.e., $\vec{n}' = \vec{n}$, then the computation is reduced to the pure-state case, where each qubit is rotated and measured with respect to a single basis pair.  As a consequence, $ |\langle\Psi_0^{\vec{n}}| \mathcal{U} |\Psi_0^{\vec{n}}\rangle|^2 = \mathcal{A}$.  On the other hand, if the projective measurement basis is different, i.e., $\vec{n}' \neq \vec{n}$, then $ |\langle\Psi_0^{\vec{n}'}| \mathcal{U} |\Psi_0^{\vec{n}}\rangle|^2 = 0$ due to the orthogonality condition $\langle\psi_i^{(m)}|\psi_j^{(n)}\rangle=\delta_{mn}\delta_{ij}$.  Combining both results, we have
\begin{equation}
 |\langle\Psi_0^{\vec{n}'}| \mathcal{U} |\Psi_0^{\vec{n}}\rangle|^2 = \mathcal{A} \delta_{\vec{n} \vec{n}'}~.
\end{equation}

The mixed-state computational task aims to obtain the  $\bigotimes_{k=1}^K(\hat{\mathbb{I}}_{L_k}+\hat{Z}_{L_k})=\sum_{\vec{n}}|\Psi_0^{\vec{n}} \rangle\langle \Psi_0^{\vec{n}}|$ measurement probability of the transformed state $\mathcal{U} \rho_0 \mathcal{U}^\dag$.  The result is given by $\textrm{Tr}\big\{\mathcal{U} \rho_0 \mathcal{U}^\dag (\sum_{\vec{n}}|\Psi_0^{\vec{n}} \rangle\langle \Psi_0^{\vec{n}}|) \big\}=\sum_{\vec{n}} p_{\vec{n}} |\langle\Psi_0^{\vec{n}}| \mathcal{U} |\Psi_0^{\vec{n}}\rangle|^2 = \sum_{\vec{n}} p_{\vec{n}}\mathcal{A} = \mathcal{A}$, which is identical to the pure-state case.

\section{Appendix II: Exponential-Parity Gate}

We here show the details of implementing exponential-parity gates in Eq.~(\ref{eq:gate}), while the implementation of exponential-swap can be found in Refs.~\cite{Lau:2016QML, Lau:2016UUQC, Iman:2016UQE}.  We first note that the single-mode and two-mode exponential-parity gates can be expressed as
\begin{equation}
e^{i\theta \hat{\mathcal{P}}}=\cos\theta \hat{\mathbb{I}} +i \sin\theta \hat{\mathcal{P}}~,~e^{i\theta \hat{\mathcal{P}}\otimes\hat{\mathcal{P}} }=\cos\theta \hat{\mathbb{I}} +i \sin\theta \hat{\mathcal{P}}\otimes\hat{\mathcal{P}} \nonumber
\end{equation}
Consider the auxiliary qubit is prepared as $|+\rangle$ and a single-qumode state $|\psi\rangle$.  The first operation in Eq.~(\ref{eq:gate}) is 
\begin{eqnarray}
|+\rangle|\psi\rangle &\overset{\hat{\mathcal{C}}}{\longrightarrow}& \frac{1}{\sqrt{2}}|0\rangle|\psi\rangle + \frac{1}{\sqrt{2}}|1\rangle\hat{\mathcal{P}}|\psi\rangle \nonumber \\
&\overset{\hat{R}_X(\theta)}{\longrightarrow}& \frac{1}{\sqrt{2}}(\cos\theta|0\rangle+i\sin\theta|1\rangle)|\psi\rangle \nonumber \\
&&+ \frac{1}{\sqrt{2}}(i\sin\theta|0\rangle+\cos\theta|1\rangle)\hat{\mathcal{P}}|\psi\rangle \nonumber \\
&\overset{\hat{\mathcal{C}}}{\longrightarrow}& |+\rangle (\cos\theta|\psi\rangle+i\sin\theta \hat{\mathcal{P}}|\psi\rangle) =|+\rangle e^{i\theta \hat{\mathcal{P}}}|\psi\rangle~.\nonumber 
\end{eqnarray}
Similarly for the third operation in Eq.~(\ref{eq:gate}), any two-qumode state $|\psi_1\rangle |\psi_2\rangle$ transforms as
\begin{eqnarray}
|+\rangle|\psi_1\rangle|\psi_2\rangle &\overset{\hat{\mathcal{C}}_1\hat{\mathcal{C}}_2}{\longrightarrow}& \frac{1}{\sqrt{2}}|0\rangle|\psi_1\rangle|\psi_2\rangle + \frac{1}{\sqrt{2}}|1\rangle\hat{\mathcal{P}}|\psi_1\rangle \hat{\mathcal{P}}|\psi_2\rangle \nonumber \\
&\overset{\hat{R}_X(\theta)}{\longrightarrow}& \frac{1}{\sqrt{2}}(\cos\theta|0\rangle+i\sin\theta|1\rangle)|\psi_1\rangle|\psi_2\rangle \nonumber \\
&&+ \frac{1}{\sqrt{2}}(i\sin\theta|0\rangle+\cos\theta|1\rangle)\hat{\mathcal{P}}|\psi_1\rangle \hat{\mathcal{P}}|\psi_2\rangle \nonumber \\
&\overset{\hat{\mathcal{C}}}{\longrightarrow}& |+\rangle (\cos\theta|\psi_1\rangle|\psi_2\rangle +i\sin\theta \hat{\mathcal{P}}|\psi_1\rangle \hat{\mathcal{P}}|\psi_2\rangle) \nonumber \\
&=&|+\rangle e^{i\theta \hat{\mathcal{P}}\otimes \hat{\mathcal{P}}}|\psi_1\rangle|\psi_2\rangle~.\nonumber 
\end{eqnarray}

We note that this method can be generalised to deterministically implement $e^{i \theta \hat{O}}$ gate, provided that controlled-$\hat{O}$ gate can be implemented and $\hat{O}^2=\hat{\mathbb{I}}$.

\section{Appendix III: Interaction Engineering}
 
Consider an auxiliary qubit and a qumode are evolving freely under the hybrid Hamiltonian
\begin{equation}\label{eq:H_hybrid}
H=\nu\hat{a}^\dag \hat{a} + \nu \eta \hat{Z}_A (\hat{a}+\hat{a}^\dag)~,
\end{equation}
where $\hat{Z}_A \equiv |+1\rangle_A\langle+1|_A-|-1\rangle_A\langle-1|_A$; $|\pm1\rangle_A$ are two orthogonal states of the auxiliary qubit.  
If the auxiliary qubit is in one of the $\hat{Z}_A$ eigenstates, 
it is easy to check that $e^{-i H t} |\pm1\rangle_A |\psi(0)\rangle = |\pm1\rangle_A (\hat{U}_\pm (t) |\psi(0)\rangle)$, where 
\begin{equation}
\hat{U}_\pm(t)=e^{i\eta^2 (\nu t -\sin \nu t)}e^{-i \nu t \hat{a}^\dag \hat{a}} \hat{D}(\mp\eta  (e^{i \nu t}-1))~.
\end{equation}
At $t=\pi/\nu$, the free evolution operator becomes
\begin{equation}
e^{-i \frac{\pi}{\nu} H }=e^{-i \pi \hat{a}^\dag\hat{a}}\hat{D}(\hat{Z}_A 2\eta)~,
\end{equation}
where we have neglected an unimportant global phase.

For simplicity, we assume the auxiliary qubit can be rotated much faster than the qumode evolution.  The qubit dependence of the evolution operator can be altered by rotating the auxiliary qubit before and afterwards, i.e., 
\begin{equation}
\hat{R}_\sigma\big(\pm\frac{\pi}{4}\big) e^{-i \frac{\pi}{\nu} H } \hat{R}_\sigma\big(\mp\frac{\pi}{4}\big) = 
e^{-i \pi \hat{a}^\dag\hat{a}}\hat{D}\big(\pm\frac{i}{2}[\hat{\sigma}_A,\hat{Z}_A] 2\eta\big)~,
\end{equation}
where $\hat{\sigma}_A$ can be $\hat{X}_A$ or $\hat{Y}_A$.  The operation in Eq.~(\ref{eq:H2_engineer}) can be implemented by the sequence 
\begin{widetext}
\begin{equation}
\left(e^{-i\frac{\pi}{2}\hat{a}^\dag\hat{a}}\hat{R}_Y\big(-\frac{\pi}{4}\big)e^{-i \frac{\pi}{\nu} H }\hat{R}_Y\big(\frac{\pi}{4}\big) \hat{R}_X\big(-\frac{\pi}{4}\big)e^{-i \frac{\pi}{\nu} H }\hat{R}_X\big(\frac{\pi}{4}\big)
\hat{R}_Y\big(\frac{\pi}{4}\big)e^{-i \frac{\pi}{\nu} H }\hat{R}_Y\big(-\frac{\pi}{4}\big) \hat{R}_X\big(\frac{\pi}{4}\big)e^{-i \frac{\pi}{\nu} H } \hat{R}_X\big(-\frac{\pi}{4}\big) \right)^4~,
\end{equation}
\end{widetext}
which involves free evolutions under Eq.~(\ref{eq:H_hybrid}), rapid rotations of the auxiliary qubit, and `waiting periods' that the system evolves only under the bare Hamiltonian, $H_0=\nu\hat{a}^\dag\hat{a}$.  If the hybrid interaction strength is controllable, then the waiting period is a free evolution with $\eta=0$.  Otherwise, the hybrid interaction can be cancelled by flipping the qubit frequently, i.e., for a short time $\delta t$,
\begin{equation}
\hat{R}_X(\frac{\pi}{2}) e^{-i \delta t H} \hat{R}_X(-\frac{\pi}{2})e^{-i \delta t H} \approx e^{-i 2 \delta t H_0} +O(\delta t^2)~.
\end{equation}

\section{Appendix IV: Initialisation error}

We consider the qumode is interacting with a background with excitation $N_\textrm{th}$.  The total state of the qumode and auxiliary qubit, $\rho$, evolves according to the master equation
\begin{equation} \label{eq:master_eq}
\dot{\rho} = -i [H,\rho ]+\frac{\nu}{Q} (N_\textrm{th}+1)\mathcal{D}\{ \hat{a}\}\rho+\frac{\nu}{Q} N_\textrm{th}\mathcal{D}\{ \hat{a}^\dag\}\rho~,
\end{equation}
where $H$ is the system Hamiltonian; the dissipation superoperator is defined as $\mathcal{D}\{ \hat{O}\}\rho \equiv \hat{O}\rho\hat{O}^\dag-\frac{1}{2}\hat{O}^\dag\hat{O}\rho-\frac{1}{2}\rho\hat{O}^\dag\hat{O}$.  The master equation can be rewritten in the quantum jump formalism \cite{Dalibard:1992ub, Plenio:1998jk}
\begin{eqnarray}
\dot{\rho} &=& -i (H_\textrm{eff}\rho-\rho H_\textrm{eff}^\dag)  
+  \frac{\nu}{Q} (N_\textrm{th}+1)\langle n \rangle \Big( \frac{\hat{a}\rho \hat{a}^\dag}{\langle n \rangle} \Big)
\nonumber\\
& &+\frac{\nu}{Q} N_\textrm{th} (\langle n \rangle +1) \Big( \frac{\hat{a}^\dag \rho \hat{a}}{\langle n \rangle+1} \Big)~, 
\end{eqnarray}
where the non-Hermitian Hamiltonian $H_\textrm{eff}=H-i\frac{1}{2}\frac{\nu}{Q} (N_\textrm{th}+1)\hat{a}^\dag\hat{a}-i\frac{1}{2}\frac{\nu}{Q} N_\textrm{th} \hat{a}\hat{a}^\dag$, and $\langle n \rangle = \textrm{Tr}(\hat{a}^\dag \hat{a}\rho)$.  For a short time $dt$, the probability of quantum jump is $(2 N_\textrm{th} \langle n\rangle +N_\textrm{th}+\langle n \rangle)\frac{\nu}{Q}dt$.  In the pessimistic scenario, we regard a quantum jump must induce logical error in initialising a TQP qubit.  The total initialisation error is thus the total probability of quantum jump during the implementation of $\hat{\mathcal{C}}$ in the parity measurement (\ref{eq:measure}).  

The sequence (\ref{eq:H2_engineer}) approximates the second order hybrid interaction, i.e., $H \approx H_2$.  The total duration of each sequence (\ref{eq:H2_engineer}) lasts for $t=18\pi/\nu$, so the effective $H_2$ interaction strength is $\lambda=\frac{32}{9}\eta^2 \nu$.  The total implementation time of $\hat{\mathcal{C}}$ is $t=\frac{\pi}{2 \lambda} = \frac{9 \pi}{64 \eta^2 \nu}$, so the total probability of quantum jump is
\begin{equation}
\epsilon_\textrm{TQP} \approx (2 N_\textrm{th} \langle n\rangle +N_\textrm{th}+\langle n \rangle) \frac{9 \pi}{64 \eta^2 Q} \sim \frac{N_\textrm{th} }{\eta^2 Q} (2 \langle n\rangle +1)~,
\end{equation}
for $N_\textrm{th} \gg \langle n \rangle \gtrsim 1$.

On the other hand, the system Hamiltonian for performing dissipative cooling is
\begin{equation}
H_\textrm{cool} = \frac{\Delta}{2} \hat{Z}_A +\frac{\Omega}{2} \hat{X}_A + \nu\hat{a}^\dag \hat{a}+H_1~,
\end{equation}
where $\Delta$ and $\Omega$ are detuning and Rabi frequency of the field that drives the auxiliary qubit.  The auxiliary qubit is also engineered to be dissipative \cite{Marzoli:1994tz}, i.e., two extra terms are added to the master equation (\ref{eq:master_eq})
\begin{equation}
\Gamma_\textrm{dc} \mathcal{D} \{|g\rangle \langle e|\}\rho + 2 \Gamma_\textrm{dp} \mathcal{D} \{|e\rangle \langle e|\}\rho~,
\end{equation}
where $|g\rangle$ and $|e\rangle$ are the ground state and excited state of the auxiliary qubit.  

Because the pure-state qubits are initialised from the ground state, the initialisation error is closely related to the occupation of the ground state.  Particularly, $\epsilon_\textrm{cool} \approx \langle n\rangle_f$, when the final excitation after cooling $\langle n \rangle_f \ll 1$.  When the optimal system parameters are available, the maximum cooling rate is $\tilde{\Gamma}_c \sim \eta^2 \nu^2/\Gamma_\textrm{dc}$, where the decay rate satisfies $\Gamma_\textrm{dc}\ll \nu$ in the resolved-sideband regime \cite{Rabl:2010cm}.  The final excitation is given by the equilibrium of heating and cooling rate, i.e.,
\begin{equation}
\epsilon_\textrm{cool} \sim \frac{N_\textrm{th}}{\eta^2 Q} \frac{\Gamma_\textrm{dc}}{\nu}~.
\end{equation}

Although $\epsilon_\textrm{TQP}>\epsilon_\textrm{cool}$ in the resolved-sideband regime if TQP encoding is implemented by the sequence (\ref{eq:H2_engineer}), obtaining the scaling of optimal cooling rate requires the dephasing rate is at most at the order of the decay rate, i.e., $\Gamma_\textrm{dp} \lesssim \Gamma_\textrm{dc}$.  In some situations, the implementation of cooling could induce a large dephasing rate that the sideband is not resolved, i.e., $\Gamma_\textrm{dc}\ll \nu \ll \Gamma_\textrm{dp}$.  By using standard analysis of dissipative cooling \cite{Cirac:1992tp,Jaehne:2008gw, Rabl:2010cm, Lau:2016cool}, and considering only the leading order terms, we obtain the cooling rate
\begin{equation}
\tilde{\Gamma}_c \approx \frac{4 \eta^2 \nu^2 \Gamma_\textrm{dc} \Gamma_\textrm{dp} \Delta \Omega^2}{\nu (\Gamma_\textrm{dp}^2 + \Delta^2+\Omega^2) (\Gamma_\textrm{dc}(\Gamma_\textrm{dp}^2+\Delta^2)+\Gamma_\textrm{dp}\Omega^2)}~.
\end{equation}
After optimising the system parameters, we find that the maximum cooling rate is $\tilde{\Gamma}_c\sim \eta^2 \nu\Gamma_\textrm{dc}/\Gamma_\textrm{dp}$, so the final excitation of cooling is
\begin{equation}
\epsilon_\textrm{cool} \sim \frac{N_\textrm{th}}{\eta^2 Q} \frac{\Gamma_\textrm{dp}}{\Gamma_\textrm{dc}}~.
\end{equation}

We note that the large $\Gamma_\textrm{dp}$ might not significantly affect $\epsilon_\textrm{TQP}$.  It is because in the sequence (\ref{eq:H2_engineer}) the auxiliary qubit is dephased only when it is rotated, but not during the free evolution.  
The dephasing error can thus be suppressed by reducing the rotation time, i.e., using a large Rabi frequency.

\section{Appendix V: Noiseless Subsystems}

If a pure-state encoding state $|\psi\rangle$ is resilient to both the two-qumode collective phase noise and two-qumode collective squeezing noise, it must satisfy
\begin{equation}
\hat{\mathcal{E}}_P(\phi)|\psi\rangle = e^{i \mu_P(\phi)}|\psi\rangle~\textrm{ and }~\hat{\mathcal{E}}_S(\xi)|\psi\rangle = e^{i \mu_S(\xi)}|\psi\rangle~,
\end{equation}
where $\phi$ and $\xi$ are respectively the phase-shift noise and squeezing noise parameters ; $\mu_P(\phi)$ and $\mu_S(\xi)$ are global phases that do not decohere the quantum information.  Differentiating both equations with respect to the noise parameters, we have 
\begin{eqnarray}\label{eq:phase_eigen}
(\hat{a}^\dag_1\hat{a}_1+\hat{a}^\dag_2\hat{a}_2)|\psi\rangle &=& (\partial_\phi \mu_P(\phi))|\psi\rangle~, \\
(\hat{a}_1^2 -\hat{a}^{\dag 2}_1+\hat{a}_2^2 -\hat{a}^{\dag 2}_2)|\psi\rangle &=& (i\partial_\xi \mu_S(\xi))|\psi\rangle~, \label{eq:squeeze_eigen}
\end{eqnarray}
where we have assumed $|\psi\rangle$ is encoded by qumodes $1$ and $2$.  Because the qumode operators and $|\psi\rangle$ are independent of the noise parameters, $\partial_\phi \mu_P(\phi)$ and $i\partial_\xi \mu_S(\xi)$ are independent of $\phi$ and $\xi$. Then Eqs.~(\ref{eq:phase_eigen}) and (\ref{eq:squeeze_eigen}) imply that $|\psi\rangle$ has to be a simultaneous eigenstate of $\hat{a}^\dag_1\hat{a}_1+\hat{a}^\dag_2\hat{a}_2$ and $\hat{a}_1^2 -\hat{a}^{\dag 2}_1+\hat{a}_2^2 -\hat{a}^{\dag 2}_2$.

$\hat{a}^\dag_1\hat{a}_1+\hat{a}^\dag_2\hat{a}_2$ is the total excitation number operator of the two qumodes.  Its eigenvalues and eigenstates can be easily found as $M$ and $\{|M-m\rangle_1 |m\rangle_2 \}$, where $M$ is any non-negative integer and $m=0,\ldots M$.  These eigenstates constitute the decoherence-free subspaces of the collective phase-shift noise \cite{Lidar:1998wl}.  Any state satisfying Eq.~(\ref{eq:phase_eigen}) must be expressed as $|\psi\rangle = \sum_{m=0}^M c_m |M-m\rangle_1|m\rangle_2$ for a fixed $M$.  However, Eq.~(\ref{eq:squeeze_eigen}) is violated for non-zero $i\partial_\xi \mu_S(\xi)$, because the qumode operators on the left hand side do not preserve the total excitation number.  

The only remaining situation is $i\partial_\xi \mu_S(\xi)=0$, i.e.,
\begin{equation}\label{eq:eigen_zero}
(\hat{a}_1^2 -\hat{a}^{\dag 2}_1+\hat{a}_2^2 -\hat{a}^{\dag 2}_2) \sum_{m=0}^M c_m |M-m\rangle_1|m\rangle_2 =0~.
\end{equation}
By considering the components $|M+2\rangle_1|0\rangle_2$ and $|0\rangle_1|M+2\rangle_2$, we can deduce $c_0=c_N=0$.  Then Eq.~(\ref{eq:eigen_zero}) is reduced to
\begin{equation}
(\hat{a}_1^2 -\hat{a}^{\dag 2}_1+\hat{a}_2^2 -\hat{a}^{\dag 2}_2) \sum_{m=1}^{M-1} c_m |M-m\rangle_1|m\rangle_2 =0~.
\end{equation}
By induction, it can be easily shown that $c_m=0$ for all $m$.  As a result, there does not exist a pure-state encoding state $|\psi\rangle$ that is resilient to both the collective phase-shift noise and the collective squeezing noise.

\end{document}